\documentclass[aps,twocolumn,longbibliography,nofootinbib]{revtex4-1}
\usepackage[utf8]{inputenc}
\usepackage[T1]{fontenc}
\usepackage{amsmath}
\usepackage{amssymb}
\usepackage{amsthm}
\usepackage{siunitx}
\usepackage{physics}
\usepackage{graphicx}
\usepackage{subfigure}
\usepackage{tabularx}
\usepackage{booktabs}
\usepackage{multirow}
\usepackage{hyperref}

\hypersetup{hypertex=true,
colorlinks=true,
linkcolor=blue,
urlcolor = blue,
citecolor=blue}

\begin{document}

\title{Topological bulk and edge correlations of BCS condensate in a
two-dimensional singlet-triplet spin pairing model}
\author{E. S. Ma}
\author{K. L. Zhang}
\email{zkl@mail.nankai.edu.cn}
\author{Z. Song}
\email{songtc@nankai.edu.cn}
\affiliation{School of Physics, Nankai University, Tianjin 300071, China}

\begin{abstract}
The condensate of the Bardeen-Cooper-Schrieffer (BCS) pair in the ground
state, which may contain information on both topology and spin pairing,
promises the superconductivity of the system. In this paper, we study a
singlet-triplet spin paring model on a square lattice and investigate the
consequences of the competition of on-site and nearest neighbor pairing
parameters. We show that the ground state of the system has the form of the
condensate of the BCS pair, and the topological transition is associated with
the nonanalytic behavior of the pairing order parameters. A real space
correlation function on opposite spin direction is introduced to
characterizing the topological phase of the many-body ground state.
Numerical results demonstrate that this method works well in the presence of
disordered perturbation, lattice defects, or irregular boundary conditions.
The real space correlation function between two edges of the system is also
discussed, which directly reflects the existence of topological edge modes
in the many-body ground state.
\end{abstract}

\maketitle

\section{Introduction}

The topological phase of matter has received much attention in recent decades
due to its robust physical properties, which offer potential applications
for novel devices and quantum information technology \cite{Klitzing1980,
Laughlin1981, Thouless1982, Laughlin1983, Haldane1988, Kane2005, Konig2007,
Fu2007a, Qi2009, qi2011topological, Chang2013, wong2013majorana, Chiu2014,
Liu2017, Zhang2019, xie2023antihelical}. From the perspective of topological
band theory, these phases fall into two categories \cite%
{chiu2016classification}: fully gapped topological phases, such as topological
insulators and topological superconductors \cite{Schnyder2008, Qi2009,
qi2011topological}, and gapless topological phases, such as topological semimetals
and nodal superconductors \cite{beri2010topologically, wong2013majorana,
Chiu2014, queiroz2014stability, schnyder2015topological,
bouhon2018topological, kobayashi2018symmetry, nayak2021evidence,
xie2021time, bazarnik2023antiferromagnetism}. The common characteristic of
these topological matters is the existence of topological protected edge
modes. Unlike topological insulators (semimetals), the edge modes of
topological (nodal) superconductors neither particles nor holes but
Bogoliubov quasiparticles, which provide a superconducting channel at the
boundary. In terms of the classification of superconducting pairing about
spin structures, the Cooper pairs may contain singlet or triplet pairing
components, which is frequently discussed in the realm of superconductivity 
\cite{mineev1999introduction, gor2001superconducting, aperis2008coexistence,
bergeret2013singlet, wang2022singlet}.


The existence of topological protected edge modes can be predicted by the
bulk topological invariants constructed from the bulk Hamiltonian in
momentum space, which is referred to as bulk-boundary correspondence (BBC) 
\cite{hatsugai1993chern, kellendonk2002edge, qi2006general, mong2011edge,
essin2011bulk} and is of fundamental importance in studies of the
topological phase of matter. In recent years, more efforts have been made to
develop real-space characterization methods for topological
phases, for example,  real-space topological markers constructed from
projectors and position operators \cite{bianco2011mapping, sykes2021local,
chen2023optical, chen2023universal} and approaches based on correlation
functions \cite{ringel2011determining, lepori2023strange} or entanglement
spectra \cite{li2008entanglement, turner2010entanglement,
prodan2010entanglement}. The main advantage of these methods is that they
are more relevant to real systems in experiments, i.e., for systems
without translation symmetry, examples of which include systems with
defects or disorder \cite{bianco2011mapping, chen2023universal}. The topological
Anderson insulator predicted and discovered in recent years \cite{Li2009,
Zhang2012, Meier2018} is another well-known example for the role of 
real-space characterization methods. Most recently, theoretical proposal
suggests that some topological markers may be measured by real-space
experiments \cite{chen2023optical}.

In this paper, we study a mixed singlet-triplet spin pairing model on a square
lattice, the Hamiltonian of which is quadratic and includes pairing
terms, i.e., on-site pairing and nearest neighbor pairing between
opposite spin directions. We investigate the consequences of the competition
of different pairing parameters. It shows that the ground state of the
system has the form of the condensate of the Bardeen-Cooper-Schrieffer (BCS) \cite%
{Bardeen1957} pair. To characterize the phase transition and the properties
of the ground state, we introduce the pairing order parameter and find that
the topological transition is associated with the nonanalytic behavior of
the order parameter. Furthermore, we find that the phase factor in the
ground state related to the system topology can be revealed by a real space
correlation function in the opposite spin direction, which provides a real-space
scheme for detecting the topological phase of a class of systems. Numerical
results demonstrate that this method works well in the presence of
disordered perturbation, random lattice defects, or when irregular boundary
conditions are adopted. Besides, we also compute the real space correlation
function between two edges of the system, which directly reflects the
existence of topological edge modes in the many-body ground state, verifying
the BBC from the perspective of real space bulk and edge correlation functions.

This paper is organized as follows. In Sec. \ref{Model}, we introduce the
model, reveal the topological phase diagram, and show that the ground
state has the form of the condensate of the BCS pair. In Sec. \ref%
{BBCCorrelation}, we investigate the real space bulk and edge correlation
functions of the many-body ground state of the system. Finally, we summarize
and discuss the results of the paper in Sec. \ref{Summary}.

\section{Model and phase diagram}

\label{Model}

First, we consider a mixed singlet-triplet spin paring model defined on a
square lattice, the Hamiltonian of which has the form 
\begin{eqnarray}
H &=&\sum_{\mathbf{r}}\sum_{\mathbf{a}=\hat{x},\hat{y}}(\Delta _{+}c_{%
\mathbf{r,\downarrow }}c_{\mathbf{r\mathbf{+}a,\uparrow }}+\Delta _{-}c_{%
\mathbf{r+a,\downarrow }}c_{\mathbf{r,\uparrow }})  \notag \\
&&+\Delta _{0}\sum_{\mathbf{r}}c_{\mathbf{r,\downarrow }}c_{\mathbf{%
r,\uparrow }}+\mathrm{H.c.},  \label{H_real_space}
\end{eqnarray}%
where $\Delta _{0}$ and $\Delta _{\pm }$ are real parameters for on-site and
nearest neighbor pairings, respectively. The index $\mathbf{r}=(m,n)$
denotes the lattice coordinate; $\hat{x}$ and $\hat{y}$ are unit vectors in
the $x$ and $y$ directions. Different from most related works, the
Hamiltonian in Eq. (\ref{H_real_space}) only contains the pairing terms, and
we are interested in the consequence of the competition between the on-site
and nearest neighbor pairings.

Employing the periodic boundary conditions in both directions and applying
the Fourier transformation 
\begin{equation}
c_{\mathbf{k,}\sigma }=\sum_{\mathbf{r}}e^{i\mathbf{k\cdot r}}c_{\mathbf{r,}%
\sigma },
\end{equation}%
we obtain the Hamiltonian in $\mathbf{k}$ space 
\begin{equation}
H=\sum_{\mathbf{k}}C_{\mathbf{k}}^{\dagger }H_{\mathrm{BdG}}\left( \mathbf{k}%
\right) C_{\mathbf{k}},  \label{H_k}
\end{equation}%
where the Numbu spinor is defined as $C_{\mathbf{k}}^{\dagger }=\left( 
\begin{array}{cccc}
c_{\mathbf{k,\uparrow }}^{\dagger } & c_{-\mathbf{k,\downarrow }} & c_{%
\mathbf{k,\downarrow }}^{\dagger } & c_{-\mathbf{k,\uparrow }}%
\end{array}%
\right) $. The Bogoliubov-de-Gennes (BdG) representation of the Hamiltonian
is a block diagonal matrix: 
\begin{equation}
H_{\mathrm{BdG}}\left( \mathbf{k}\right) =\frac{1}{2}\left( 
\begin{array}{cc}
H\left( \mathbf{k}\right) & \mathbf{0} \\ 
\mathbf{0} & -H\left( -\mathbf{k}\right)%
\end{array}%
\right) ,  \label{H_BdG}
\end{equation}%
where $H\left( \mathbf{k}\right) $ represents a pseudo spin Hamiltonian $%
H\left( \mathbf{k}\right) =B_{x}(\mathbf{k})\sigma _{x}+B_{y}(\mathbf{k}%
)\sigma _{y}$ in the effective magnetic field 
\begin{eqnarray}
B_{x}(\mathbf{k}) &=&\left( \Delta _{+}+\Delta _{-}\right) \left( \cos
k_{x}+\cos k_{y}\right) +\Delta _{0},  \notag \\
B_{y}(\mathbf{k}) &=&\left( \Delta _{+}-\Delta _{-}\right) \left( \sin
k_{x}+\sin k_{y}\right) .
\end{eqnarray}
In fact, Hamiltonian $H\left( \mathbf{k}\right) $ can be related to a
spinless Kitaev model by the unitary transformation $c_{\mathbf{k,\uparrow }%
}=(c_{\mathbf{k}}-c_{-\mathbf{k}}^{\dagger })/\sqrt{2},c_{-\mathbf{%
k,\downarrow }}^{\dagger }=(c_{\mathbf{k}}+c_{-\mathbf{k}}^{\dagger })/\sqrt{%
2}$. In this sense, $c_{\mathbf{k,\uparrow }}$ and $c_{-\mathbf{k,\downarrow 
}}^{\dagger }$ are pseudo-spin operators.

The real functions $B_{x}(\mathbf{k})$ and $B_{y}(\mathbf{k})$ are the
coefficients of singlet and triplet pairing in momentum space. The Fermi
statistics place constraints on the forms of functions $B_{x}(\mathbf{k})$ and 
$B_{y}(\mathbf{k})$. For example, we take the singlet term, 
\begin{eqnarray}
&&\sum_{\mathbf{k}}[B_{x}(\mathbf{k})(c_{\mathbf{k,\uparrow }}^{\dagger }c_{-%
\mathbf{k,\downarrow }}^{\dagger }-c_{\mathbf{k,\downarrow }}^{\dagger }c_{-%
\mathbf{k,\uparrow }}^{\dagger })  \notag \\
&=&\sum_{\mathbf{k}}[B_{x}(-\mathbf{k})(-c_{\mathbf{k,\downarrow }}^{\dagger
}c_{-\mathbf{k,\uparrow }}^{\dagger }+c_{\mathbf{k,\uparrow }}^{\dagger }c_{-%
\mathbf{k,\downarrow }}^{\dagger })\text{,}
\end{eqnarray}%
so that $B_{x}(\mathbf{k})=B_{x}(-\mathbf{k})$. Similarly, we have $B_{y}(%
\mathbf{k})=-B_{y}(-\mathbf{k})$ for the triplet term. This constraint can
also be given by the particle-hole symmetry of the BdG Hamiltonian, that is $%
H_{\mathrm{BdG}}\left( \mathbf{k}\right) =-\mathcal{C}H_{\mathrm{BdG}}\left(
-\mathbf{k}\right) \mathcal{C}^{-1}$, where $\mathcal{C}=\sigma _{x}\otimes
\sigma _{x}\mathcal{K}$ and $\mathcal{K}$ is the complex-conjugation
operator. The model also obeys time reversal and inversion symmetry, i.e.,
for the BdG Hamiltonian we have $H_{\mathrm{BdG}}\left( \mathbf{k}\right) =%
\mathcal{T}H_{\mathrm{BdG}}\left( -\mathbf{k}\right) \mathcal{T}^{-1}$ and $%
H_{\mathrm{BdG}}\left( \mathbf{k}\right) =\mathcal{P}H_{\mathrm{BdG}}\left( -%
\mathbf{k}\right) \mathcal{P}^{-1}$, with $\mathcal{T}=\mathcal{K}$ and $%
\mathcal{P}=\sigma _{0}\otimes \sigma _{x}$; $\sigma _{0}$ is a $2\times
2$ identity matrix.

\begin{figure}[t]
\centering
\includegraphics[width=0.5\textwidth]{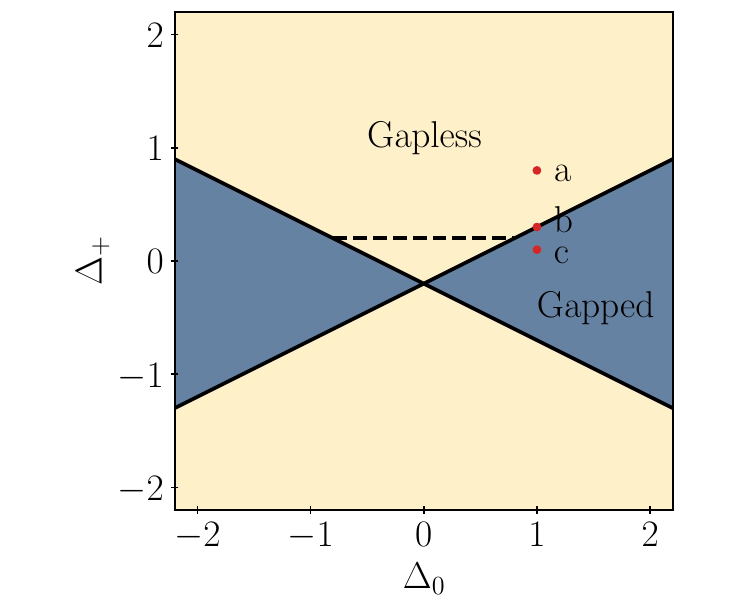}
\caption{Phase diagram in the $\Delta_{0}$-$\Delta_{+}$ parameter plane for
the model studied in this paper. The black lines indicate the phase boundary
that separates the topological trivial (gapped; gray region) and nontrivial
(gapless; yellow region) phases identified by the winding number of the
vector field $\widehat{\mathbf{B}}$ or the order parameter $O$ of the ground
state. The dashed line represents the parameter $\Delta _{+}=\Delta _{-}$,
where the system is trivial. The other parameter is set as $\Delta_{-}=0.2$. The
three red dots correspond to the parameters of the systems taken in the
numerical computations for Figs. \protect\ref{Arrows} (a)-(c).}
\label{PhaseDiagram}
\end{figure}

\begin{figure}[t]
\centering
\includegraphics[width=0.4\textwidth]{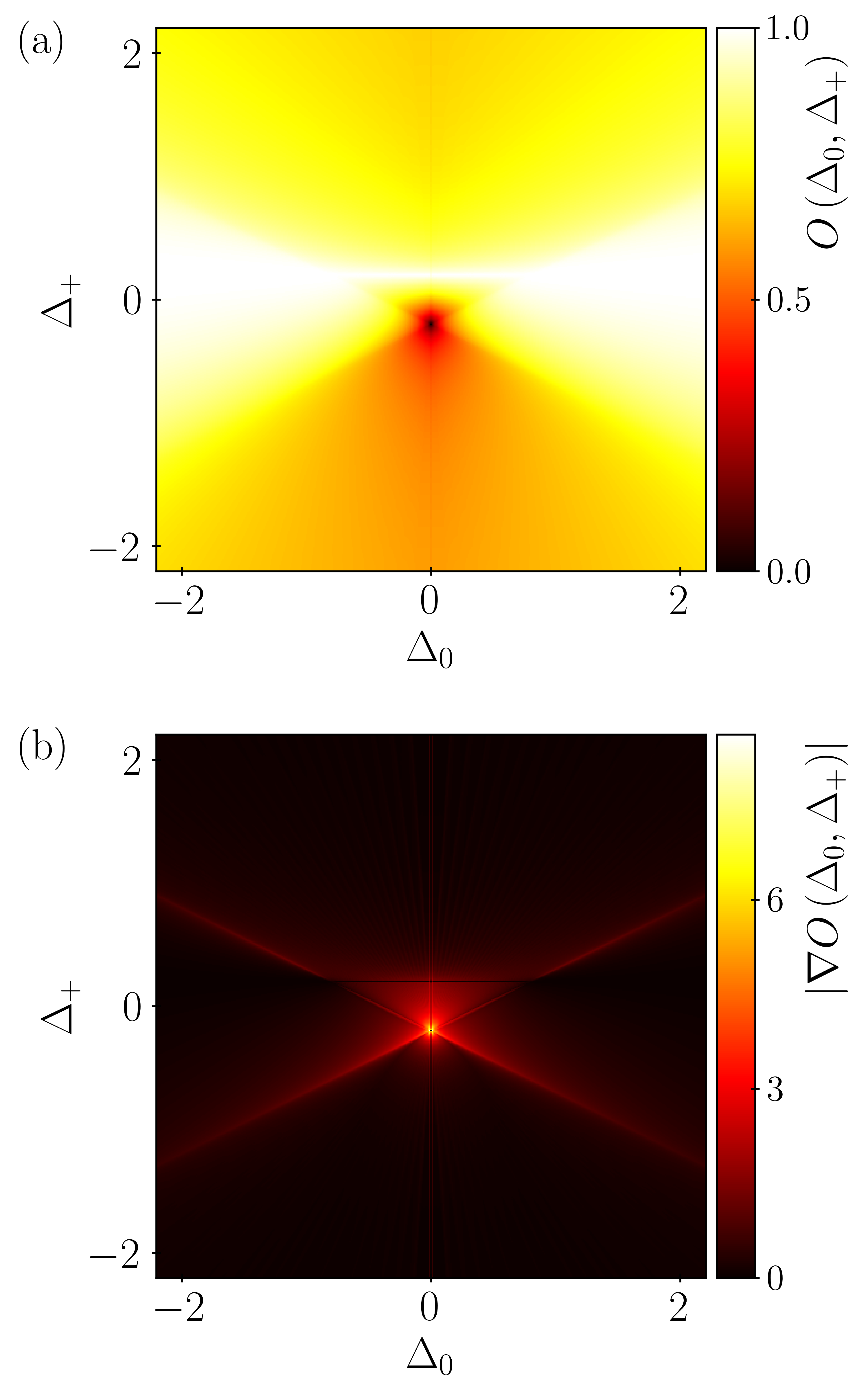}
\caption{Numerical results of the order parameter. (a) The pairing order
parameter $O$ in Eq. (\protect\ref{OrderO0}) in the $\Delta_{0}$-$\Delta_{+}$
parameter plane. (b) The corresponding absolute value of the gradient of the
order parameter $O$ in the $\Delta_{0}$-$\Delta_{+}$ plane, which indicates
the phase boundary. Other parameters are set as $\Delta_{-}=0.2$ and $N=100$%
. }
\label{OrderParameter}
\end{figure}

By diagonalizing the Hamiltonian $H$ in Eq. (\ref{H_k}), the ground state is
obtained as 
\begin{equation}
\left\vert \text{G}\right\rangle =\prod_{\mathbf{k}}\frac{1+e^{i\phi _{%
\mathbf{k}}}c_{-\mathbf{k\downarrow }}^{\dagger }c_{\mathbf{k\uparrow }%
}^{\dagger }}{\sqrt{2}}\left\vert 0\right\rangle ,  \label{G_state}
\end{equation}%
where the angle is%
\begin{equation}
\phi _{\mathbf{k}}=\arg \left( B_{x}-iB_{y}\right) .
\end{equation}%
Note that the ground state can be rewritten in the form%
\begin{equation}
\left\vert \text{G}\right\rangle =\sum_{n=0}^{N^{2}}\frac{2^{-N^{2}/2}}{n!}%
\left( s^{+}\right) ^{n}\left\vert 0\right\rangle .  \label{BCScon}
\end{equation}%
Unlike the conventional BCS wave function, the ground state in Eq. (\ref%
{BCScon}) describes the condensate of the BCS pair. The operators%
\begin{eqnarray}
s^{+} &=&\left( s^{-}\right) ^{\dag }=\sum_{\mathbf{k}}e^{i\phi _{\mathbf{k}%
}}c_{-\mathbf{k,\downarrow }}^{\dag }c_{\mathbf{k,\uparrow }}^{\dagger }, 
\notag \\
s^{z} &=&\frac{1}{2}\sum_{\mathbf{k}}\left( c_{\mathbf{k,\uparrow }%
}^{\dagger }c_{\mathbf{k,\uparrow }}+c_{-\mathbf{k,\downarrow }}^{\dag }c_{-%
\mathbf{k,\downarrow }}-1\right) ,
\end{eqnarray}%
are pseudo-spin operators that satisfy the Lie algebra commutation relation $%
\left[ s^{z},s^{\pm }\right] =\pm s^{\pm }$.

We note that the angle $\phi _{\mathbf{k}_{\mathrm{c}}}$\ is ill-defined at
the zero point of $\left\vert \mathbf{B}\right\vert $ with%
\begin{equation}
B_{x}(\mathbf{k}_{\mathrm{c}})=B_{y}(\mathbf{k}_{\mathrm{c}})=0,
\end{equation}%
which corresponds to the topological defect of the vector field $\widehat{%
\mathbf{B}}=(\cos \phi _{\mathbf{k}},\sin \phi _{\mathbf{k}})$ if the
solutions of $\mathbf{k}_{\mathrm{c}}=\left( k_{x\mathrm{c}},k_{y\mathrm{c}%
}\right) $ are isolated points in the $\mathbf{k}$-plane. In this sense,
the condensate of the collective BCS-pair state $s^{+}\left\vert
0\right\rangle $ is topologically nontrivial and is characterized by the
vortex of field $\widehat{\mathbf{B}}$. Obviously, such a ground state
is a gapless state. In fact, when $\Delta _{+}\neq\Delta _{-}$, we have 
\begin{equation}
k_{x\mathrm{c}}=-k_{y\mathrm{c}}=\pm \arccos \left[ -\frac{\Delta _{0}}{%
2\left( \Delta _{+}+\Delta _{-}\right) }\right] ,
\end{equation}%
in the topological nontrivial region $\left\vert \Delta _{0}\right\vert
<2\left\vert \Delta _{+}+\Delta _{-}\right\vert $ ($\Delta _{+}\neq\Delta
_{-}$). The two zero points $\left( k_{x\mathrm{c}},k_{y\mathrm{c}}\right) $
and $\left( -k_{x\mathrm{c}},-k_{y\mathrm{c}}\right) $ are Dirac points in
momentum space, the topological nature of which are characterized by the
winding number \cite{ryu2002topological, sun2012topological,
wong2013majorana, matsuura2013protected} of the vortex in the vector field $%
\widehat{\mathbf{B}}$. The combination of the time reversal and inversion
symmetry protects the Dirac points in the following sense: the diagonal term
in $H_{\mathrm{BdG}}(\mathbf{k})$ that openings a gap is forbidden by time
reversal and inversion symmetry. The position of the Dirac points only shifts
when changing the system parameters until the Dirac points merge and
open a gap when $\left\vert \Delta _{0}\right\vert \geqslant 2\left\vert
\Delta _{+}+\Delta _{-}\right\vert $. The phase diagram of the system is
presented in Fig. \ref{PhaseDiagram}. In the next section, we will show that
the vector field $\widehat{\mathbf{B}}$ can be extracted from the real space
correlation function of the ground state, where the periodic boundary
condition is no longer needed.

To characterize the phase transition and the properties of the ground state,
we introduce the following pairing order parameter 
\begin{equation}
O=\frac{1}{N^{2}}\sum_{\mathbf{k}}\left\vert \left\langle \text{G}%
\right\vert c_{\mathbf{k,\uparrow }}^{\dagger }c_{-\mathbf{k,\downarrow }%
}^{\dag }+c_{-\mathbf{k,\downarrow }}c_{\mathbf{k,\uparrow }}\left\vert 
\text{G}\right\rangle \right\vert .
\end{equation}%
Direct calculation shows that 
\begin{equation}
O=\frac{1}{N^{2}}\sum_{\mathbf{k}}\left\vert \cos \phi _{\mathbf{k}%
}\right\vert ,  \label{OrderO0}
\end{equation}%
which characterizes the pairing channel in $\mathbf{k}$ space, and is
related to the angle $\phi _{\mathbf{k}}$ in the vector field $\widehat{%
\mathbf{B}}$ that contains information of the system topology.

In Fig. \ref{OrderParameter} (a), we plot the numerical results of the order
parameter $O$ in the $\Delta_{0}$-$\Delta_{+}$ parameter plane. We can see
that in the gapped phase, the $\mathbf{k}$-space pairing strength is
stronger than that in the gapless phase. The absolute value of the gradient
of the order parameter $O$ presented in Fig. \ref{OrderParameter} (b)
indicates that topological phase transition is associated with the
nonanalytic behavior of the order parameter $O$.

\begin{figure*}[t]
\centering
\includegraphics[width=1\textwidth]{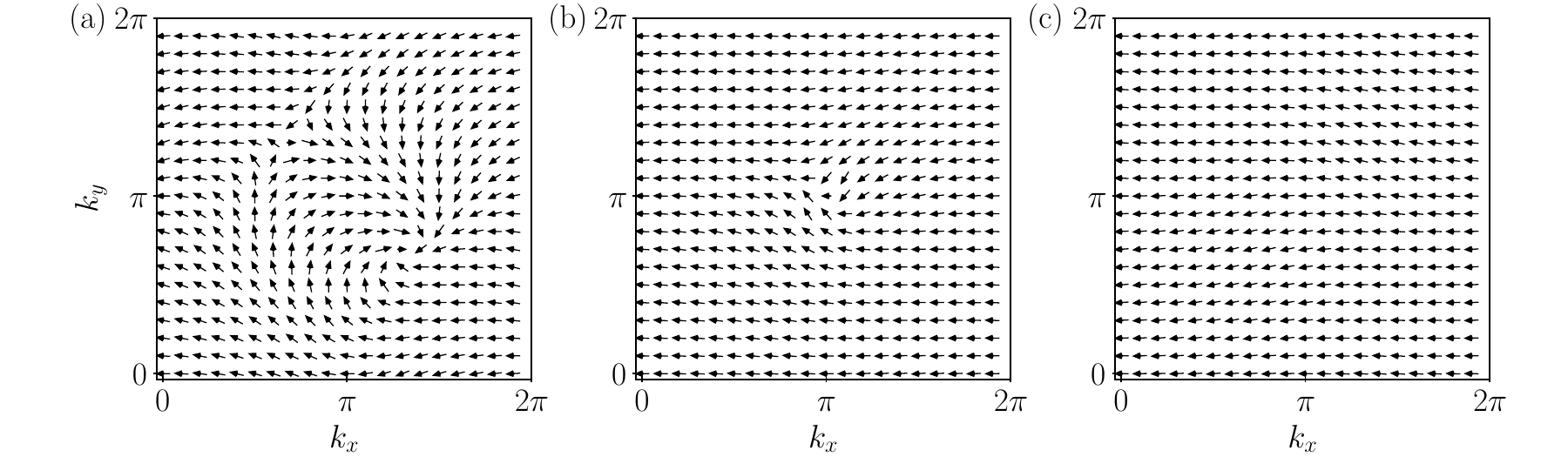}
\caption{Plots of the numerical results of the phase angle $\protect\phi _{%
\mathbf{k}}$ computed from the real space correlation function $\mathcal{C}_{%
\mathbf{r}}$ in Eq. (\protect\ref{phase_Cr}). The direction of the arrow at
different $\mathbf{k}$ represents the phase angle $\protect\phi _{\mathbf{k}%
} $. The system parameters taken are marked by the red dots in the phase
diagram of Fig. \protect\ref{PhaseDiagram}: (a) topological nontrivial case 
$\Delta_{+}=0.8$; (b) critical case $\Delta_{+}=0.3$; and (c) topological
trivial case $\Delta_{+}=0.1$. Other parameters are taken as $N=20$, $%
\Delta_{0}=1$ and $\Delta_{-}=0.2$. The open boundary condition in both
directions of the square lattice is taken.}
\label{Arrows}
\end{figure*}

\begin{figure*}[tbh]
\centering
\includegraphics[width=1\textwidth]{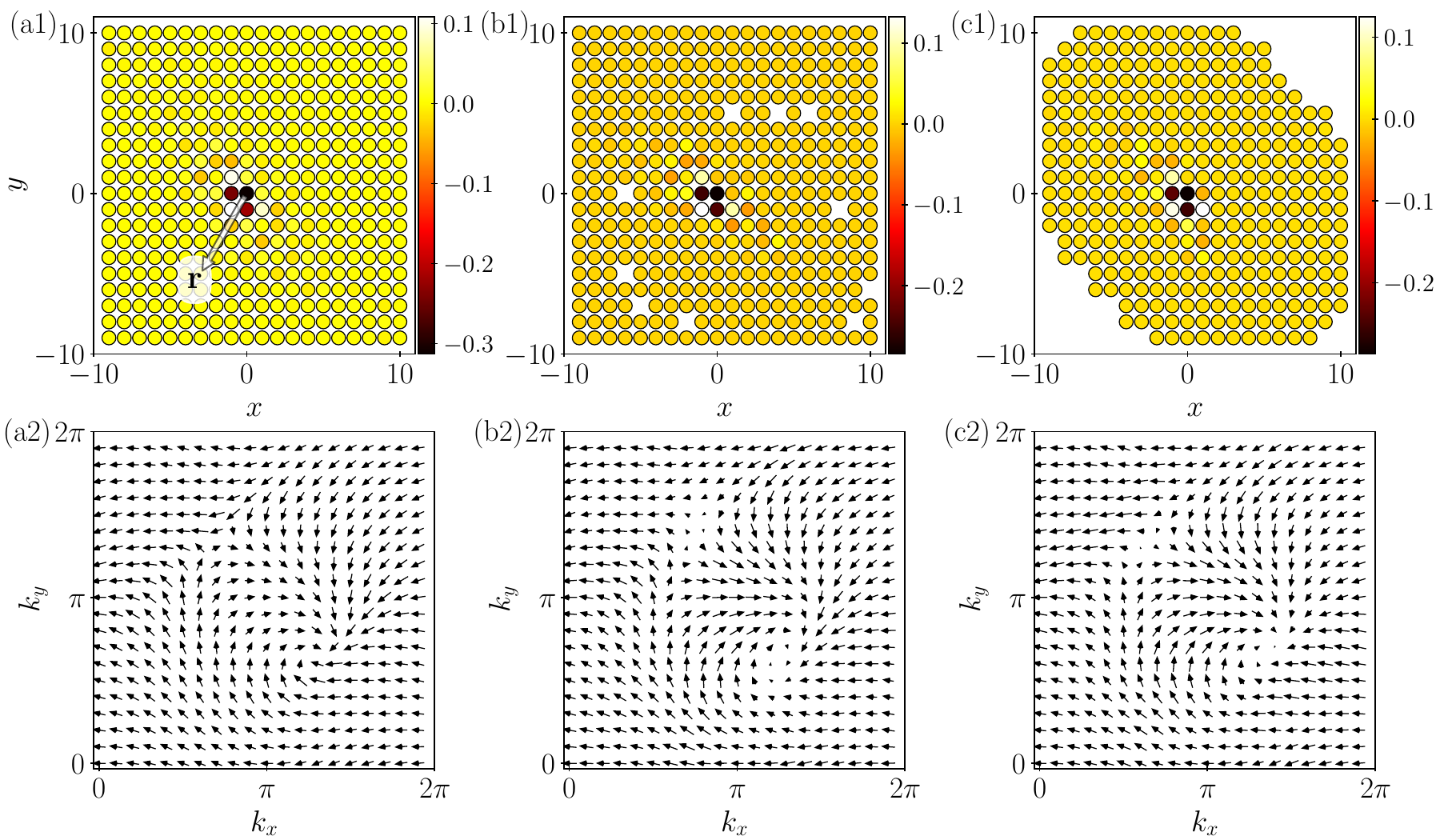}
\caption{Numerical results of the real space correlation function $\mathcal{C%
}_{\mathbf{r}}$ and phase angle $\protect\phi _{\mathbf{k}}$ in the presence
of (a) disordered perturbation, where the system parameters are nonuniform
in space and each deviates a uniform distributed random real number within
the interval $[-0.2,0.2]$; (b) random defects, where the spatial coordinates
of $10$ lattice defects are randomly taken; and (c) irregular boundary as
shown in (c1). The system parameters are taken as $\Delta_{+}=0.8$, $%
\Delta_{-}=0.2$ and $\Delta_{0}=1$.}
\label{CorrelationsArrows}
\end{figure*}

\begin{figure}[tbh]
\centering
\includegraphics[width=0.48\textwidth]{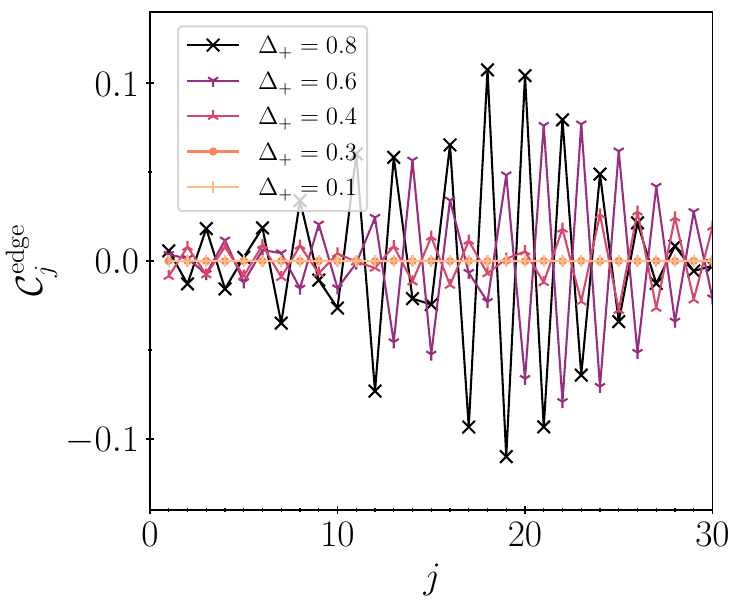}
\caption{Numerical results of the edge correlation function defined in Eq. (%
\protect\ref{edge_correlation}) for the topological nontrivial phase $%
\Delta_{+}>0.3$, critical point $\Delta_{+}=0.3$ and topological trivial
phase $\Delta_{+}=0.1$. Other system parameters are set as $N=30$, $%
\Delta_{0}=1$ and $\Delta_{-}=0.2$. The cylindrical boundary condition for the
square lattice is taken.}
\label{EdgeCorrelation}
\end{figure}

\section{Bulk-boundary correspondence by real space correlation}

\label{BBCCorrelation}

\bigskip In the previous section, we have shown that the order parameter $O$
is related to the angle $\phi _{\mathbf{k}}$ in the vector field $\widehat{%
\mathbf{B}}$, which contains information on the system topology. Therefore,
it is promising to extract the topological properties of the system from a
specific correlation function, preferably the real space correlation
function. To this end, we consider the following correlation function in
real space%
\begin{equation}
\mathcal{C}_{\mathbf{r}}=\left\langle \text{G}\right\vert c_{\mathbf{%
0,\uparrow }}c_{\mathbf{r,\downarrow }}\left\vert \text{G}\right\rangle ,
\label{Cr}
\end{equation}%
where the coordinate origin $\mathbf{0}$ is placed in the center of the
lattice when the open boundary condition is adopted. The
correlation function $\mathcal{C}_{\mathbf{r}}$\ and the phase $e^{-i\phi _{%
\mathbf{k}}}$ in the ground state are related by the Fourier transformation%
\begin{equation}
e^{i\phi _{\mathbf{k}}}=2\sum_{\mathbf{r}}e^{i\mathbf{k\cdot r}}\mathcal{C}_{%
\mathbf{r}}.  \label{phase_Cr}
\end{equation}
In fact, from the definition of the correlation function and taking the form
of the ground state $\left\vert \text{G}\right\rangle $ in Eq. (\ref{G_state}%
) into account, we have%
\begin{eqnarray}
\mathcal{C}_{\mathbf{r}} &=&\frac{1}{N^{2}}\sum_{\mathbf{k,k}^{\prime }}e^{i%
\mathbf{k}^{\prime }\cdot \mathbf{r}}\left\langle \text{G}\right\vert c_{%
\mathbf{k},\uparrow }c_{\mathbf{k}^{\prime },\downarrow }\left\vert \text{G}%
\right\rangle  \notag \\
&=&\frac{1}{N^{2}}\sum_{\mathbf{k}}e^{-i\mathbf{k}\cdot \mathbf{r}%
}\left\langle \text{G}\right\vert c_{\mathbf{k},\uparrow }c_{-\mathbf{k}%
,\downarrow }\left\vert \text{G}\right\rangle  \notag \\
&=&\frac{1}{2N^{2}}\sum_{\mathbf{k}}e^{i\phi _{\mathbf{k}}}e^{-i\mathbf{%
k\cdot r}}.
\end{eqnarray}
Thus, we have the relation in Eq. (\ref{phase_Cr}) by the Fourier
transformation. This inspires us to employ the real space correlation
function $\mathcal{C}_{\mathbf{r}}$ for detecting the topological properties
of the ground state. Different from the topological invariant defined in
momentum space, this method is applicable for systems without
translation symmetry, including systems with disorder and defects. In the
following, we present the numerical results to demonstrate our conclusions.
Numerically, the real space correlation function for the ground state of a
quadratic Hamiltonian can be computed by the method presented in 
Appendix A.

First, we compute the real space correlation function $\mathcal{C}_{%
\mathbf{r}}$ defined in Eq. (\ref{Cr}). The open boundary condition in both
directions of the lattice is taken. Although the derivation of the relation
between $\mathcal{C}_{\mathbf{r}}$ and the phase $e^{-i\phi _{\mathbf{k}}}$
in Eq. (\ref{phase_Cr}) requires periodic boundary condition, it is expected
that if $\mathcal{C}_{\mathbf{r}}$ decays rapidly with the distance between $%
\mathbf{0}$ and $\mathbf{r}$ (which is examined by the subsequent numerical
calculations), then the phase $e^{-i\phi _{\mathbf{k}}}$ is almost
unaffected by the boundary conditions. Then, we are allowed to compute the
phase $e^{-i\phi _{\mathbf{k}}}$ through Eq. (\ref{phase_Cr}) for each $%
\mathbf{k}$. We present the numerical results in Fig. \ref{Arrows}, in which
the phase angle $\phi _{\mathbf{k}}$ is denoted by the direction of the
arrow at $\mathbf{k}$. The results indicate that the phase angle can be
correctly obtained from the real space correlation function $\mathcal{C}_{%
\mathbf{r}}$: we can see that two vortices emerge in the topological
nontrivial case [Fig. \ref{Arrows} (a)] and then merge and vanish in the
critical and topological nontrivial cases [Figs. \ref{Arrows} (b) and (c)]
when system parameter $\Delta_{+}$ varies.

Furthermore, numerical simulations show that the topological feature is
robust in the presence of (a) disordered perturbation, (b) random defects
and (c) irregular boundaries. In Fig. \ref{CorrelationsArrows}, we present the
lattice geometries, the numerical results of the real space correlation
function $\mathcal{C}_{\mathbf{r}}$, and the phase angle $\phi _{\mathbf{k}}$
for these three cases. In Fig. \ref{CorrelationsArrows} (a), the system
parameters $\Delta_{+}$, $\Delta_{-}$ and $\Delta_{0}$ are nonuniform in
space, and each deviates a uniform distributed random real number within the
interval $[-0.2,0.2]$. The result in Fig. \ref{CorrelationsArrows} (a1)
indicates that the correlation function $\mathcal{C}_{\mathbf{r}}$ decays
rapidly with the distance between $\mathbf{0}$ and $\mathbf{r}$. In
comparison with Fig. \ref{Arrows} (a), the result in Fig. \ref%
{CorrelationsArrows} (a2) shows that the pattern of the vortices is robust
against disordered perturbation. Fig. \ref{CorrelationsArrows} (b) shows the
numerical results for the lattice with random defects, where the spatial
coordinates of $10$ lattice defects are randomly taken. Fig. \ref%
{CorrelationsArrows} (c) shows the numerical results for the lattice with
irregular boundaries. We can see that the signatures of the vortices are also
robust for these two cases.

Now, we turn to the investigation on the relation between the bulk topology
and edge correlation. It can be shown that in the topological nontrivial
phase, the Majorana zero modes appear at the boundaries of the system (see
Appendix B). The zero modes may contribute to the correlation function
between two edges of the system \cite{wang2017characterization,
miao2017exact}, which is one of the signatures of the system topology. To
verify this point for our model, we introduce the following edge correlation
function: 
\begin{equation}
\mathcal{C}_{j}^{\text{edge}}=\left\langle \text{G}\right\vert
c_{(1,1),\uparrow}c_{(N, j),\downarrow}\left\vert \text{G}\right\rangle,
\label{edge_correlation}
\end{equation}
where $\left\vert \text{G}\right\rangle$ is the ground state of the system
under cylindrical boundary conditions; $(1,1)$ and $(N,j)$ are the lattice
coordinates of the two ends of the system, and $j$ is the lattice coordinate
along one of the edges of the lattice cylinder. In Fig. \ref{EdgeCorrelation}%
, we present the numerical results of the edge correlation functions for the
systems in different phases with different parameter $\Delta_{+}$. This
indicates that the edge correlation function is nonzero in the topological
nontrivial phase but vanishes in the trivial phase. Therefore, we
conclude that the edge correlation function can reflect the phase diagram in
Fig. \ref{PhaseDiagram}.

The above numerical results of the bulk correlation function and the edge
correlation function in different phases verify the BBC from another
perspective, in contrast with the relation between the topological invariant
and single-particle edge mode.

\section{Summary and Discussion}

\label{Summary}

In summary, we have investigated a mixed singlet-triplet spin paring model
on a square lattice. The ground state of the system has the
form of the condensate of BCS pairs, and in the gapless phase, topological
edge modes emerge when the open boundary condition is adopted. The topological
transition is associated with the nonanalytic behavior of the order
parameter. Furthermore, we find that the phase factor in the ground state
related to the system topology can be revealed by a real space correlation
function. Numerical results demonstrate that this method works well in the
presence of disordered perturbation, lattice defects, or when irregular
boundaries condition are adopted. In addition, the results of the real space
correlation function between two edges of the system directly reflect the
existence of topological edge modes, verifying the BBC from the perspective of
real-space bulk and edge correlations.

The conclusions in this paper, including the results of numerical
simulations, reveal the consequence of the competition between the on-site
and nearest neighbor pairings in a quadratic Hamiltonian and provide
another real-space scheme for diagnosing the system topology.

\acknowledgments This work was supported by the National Natural Science
Foundation of China (under Grant No. 12374461).

\section*{Appendix A: Real space correlation}

\label{A} \setcounter{equation}{0} \renewcommand{\theequation}{A%
\arabic{equation}}

In this appendix, we present the method for computing the real space
correlation function of the ground state of the quadratic Hamiltonian. We
follow the method used in Ref. \cite{young1996numerical}. For simplicity,
the system parameters are set to be uniform in real space, and the geometry
is taken as an $N\times N$ square lattice. Other situations with disordered
perturbations, random defects and irregular boundaries can be directly
generalized.

Under the basis 
\begin{eqnarray}
C^{\dag } &=&\left( c_{\mathbf{r}_{1},\downarrow }^{\dag }\cdots c_{\mathbf{r%
}_{N^{2}},\downarrow }^{\dag },c_{\mathbf{r}_{1},\uparrow }^{\dag },\cdots
c_{\mathbf{r}_{N^{2}},\uparrow }^{\dag },\right.   \notag \\
&&\left. c_{\mathbf{r}_{1},\downarrow }\cdots c_{\mathbf{r}%
_{N^{2}},\downarrow },c_{\mathbf{r}_{1},\uparrow },\cdots c_{\mathbf{r}%
_{N^{2}},\uparrow }\right) ,
\end{eqnarray}%
the real space Hamiltonian in Eq. (\ref{H_real_space}) can be written as 
\begin{equation}
H=C^{\dag }\widetilde{H}C,
\end{equation}%
where $\widetilde{H}$ has the form 
\begin{equation}
\widetilde{H}=\left( 
\begin{array}{cc}
\mathbf{0} & M \\ 
-M & \mathbf{0}%
\end{array}%
\right) ,  \label{H_tilde}
\end{equation}%
and $M$ is a $2N^{2}\times 2N^{2}$ antisymmetric matrix. The nonzero matrix
elements are given in the following 
\begin{eqnarray}
M_{iN+j+N^{2},\left( i-1\right) N+j} &=&\frac{\Delta _{+}}{2},  \notag \\
M_{\left( i-1\right) N+j+1+N^{2},\left( i-1\right) N+j} &=&\frac{\Delta _{+}%
}{2},  \notag \\
M_{\left( i-1\right) N+j+N^{2},iN+j} &=&\frac{\Delta _{-}}{2},  \notag \\
M_{\left( i-1\right) N+j+N^{2},\left( i-1\right) N+j+1} &=&\frac{\Delta _{-}%
}{2},  \notag \\
M_{\left( i-1\right) N+j+N^{2},\left( i-1\right) N+j} &=&\frac{\Delta _{0}}{2%
},
\end{eqnarray}%
where $i,j\in \left[ 1,N-1\right] $ and each corresponding transpose matrix
element has a negative sign difference. By diagonalizing $\widetilde{H}$, we
have 
\begin{eqnarray}
H &=&C^{\dag }S\mathcal{E}S^{T}C  \notag \\
&=&\Phi ^{\dag }\mathcal{E}\Phi   \notag \\
&=&\sum_{m=1}^{N^{2}}\sum_{\sigma =\uparrow ,\downarrow }\varepsilon
_{m,\sigma }\left( \gamma _{m,\sigma }^{\dag }\gamma _{m,\sigma }-\gamma
_{m,\sigma }\gamma _{m,\sigma }^{\dag }\right) ,
\end{eqnarray}%
where $\varepsilon _{m,\sigma }\geqslant 0$ and $S$ is a real orthogonal
matrix, which has the following form 
\begin{equation}
S=\left( 
\begin{array}{cc}
\varphi  & \chi  \\ 
\chi  & \varphi 
\end{array}%
\right) ,  \label{S}
\end{equation}%
due to the particle-hole symmetry of the BdG Hamiltonian. The columns of the
matrix $S$ are formed by the eigenvectors of $\widetilde{H}$ and $\varphi $
and $\chi $ are both $2N^{2}\times 2N^{2}$ matrices. The diagonal matrix $%
\mathcal{E}$ has the form 
\begin{eqnarray}
\mathcal{E} &=&S^{T}\widetilde{H}S  \notag \\
&=&\mathrm{diag}\left( \varepsilon _{1,\downarrow },\cdots ,\varepsilon
_{N^{2},\downarrow },\varepsilon _{1,\uparrow },\cdots ,\varepsilon
_{N^{2},\uparrow },\right.   \notag \\
&&\left. -\varepsilon _{1,\downarrow },\cdots ,-\varepsilon
_{N^{2},\downarrow },-\varepsilon _{1,\uparrow },\cdots ,-\varepsilon
_{N^{2},\uparrow }\right) ,
\end{eqnarray}%
and the new basis is 
\begin{eqnarray}
\Phi ^{\dag } &=&C^{\dag }S  \notag \\
&=&\left( \gamma _{1,\downarrow }^{\dag },\cdots ,\gamma _{N^{2},\downarrow
}^{\dag },\gamma _{1,\uparrow }^{\dag },\cdots ,\gamma _{N^{2},\uparrow
}^{\dag },\right.   \notag \\
&&\left. \gamma _{1,\downarrow },\cdots ,\gamma _{N^{2},\downarrow },\gamma
_{1,\uparrow },\cdots ,\gamma _{N^{2},\uparrow }\right) ,
\label{commu_gamma}
\end{eqnarray}%
where $\left\{ \gamma _{m}^{\dag }\right\} $ are fermionic operators,
satisfying 
\begin{equation}
\left\{ \gamma _{m^{\prime },\sigma ^{\prime }},\gamma _{m,\sigma }^{\dag
}\right\} =\delta _{m^{\prime }\mathbf{,}m}\delta _{\sigma ^{\prime },\sigma
},\left\{ \gamma _{m^{\prime },\sigma ^{\prime }},\gamma _{m,\sigma
}\right\} =0.
\end{equation}

\label{B} \setcounter{figure}{0} \renewcommand{\thefigure}{B\arabic{figure}}

\begin{figure*}[tbh]
\centering
\includegraphics[width=1\textwidth]{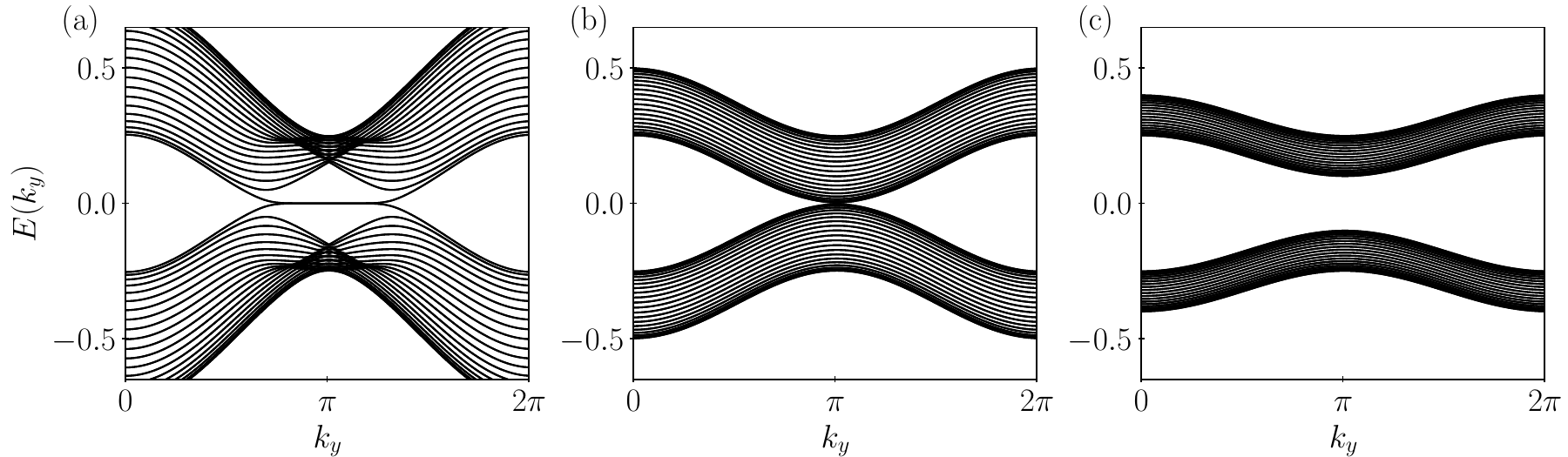}
\caption{Plots of the Majorana spectra of the Hamiltonian in Eq. (\protect
\ref{Hky}) with system size $N=20$ for three typical cases with parameters
taken the same as those in Figs. \protect\ref{Arrows} (a)-(c): (a)
topological nontrivial case $\Delta_{+}=0.8$, (b) critical case $%
\Delta_{+}=0.3$ and (c) topological trivial case $\Delta_{+}=0.1$. Other
parameters are $\Delta_{0}=1$ and $\Delta_{-}=0.2$. All the spectra are
globally twofold degenerate.}
\label{kySpectrum}
\end{figure*}

Then the real space fermionic operators can be expressed as 
\begin{eqnarray}
c_{\mathbf{r}_{i},\downarrow } &=&\sum_{m=1}^{N^{2}}\left[ \varphi
_{i,m}\gamma _{m,\downarrow }+\varphi _{i,m+N^{2}}\gamma _{m,\uparrow
}\right.   \notag \\
&&\left. +\chi _{i,m}\gamma _{m,\downarrow }^{\dag }+\chi _{i,m+N^{2}}\gamma
_{m,\uparrow }^{\dag }\right] ,
\end{eqnarray}%
and 
\begin{eqnarray}
c_{\mathbf{r}_{i},\uparrow } &=&\sum_{m=1}^{N^{2}}\left[ \varphi
_{i+N^{2},m}\gamma _{m,\downarrow }+\varphi _{i+N^{2},m+N^{2}}\gamma
_{m,\uparrow }\right.   \notag \\
&&\left. +\chi _{i+N^{2},m}\gamma _{m,\downarrow }^{\dag }+\chi
_{i+N^{2},m+N^{2}}\gamma _{m,\uparrow }^{\dag }\right] .
\end{eqnarray}%
The ground state of the system has the form 
\begin{equation}
\left\vert G\right\rangle =\prod\limits_{m=1}^{N^{2}}\prod\limits_{\sigma
=\uparrow ,\downarrow }\gamma _{m,\sigma }\left\vert 0\right\rangle .
\label{G_groundstate}
\end{equation}%
Taking the anticommutation relation of $\left\{ \gamma _{m}^{\dag }\right\} $
in Eq. (\ref{commu_gamma}) and the form of the ground state $\left\vert
G\right\rangle $ in Eq. (\ref{G_groundstate}) into account, the real space
correlation function can be computed as 
\begin{eqnarray}
&&\left\langle G\right\vert c_{\mathbf{r}_{j}\mathbf{,\uparrow }}c_{\mathbf{r%
}_{i}\mathbf{,\downarrow }}\left\vert G\right\rangle   \notag \\
&=&\sum_{m=1,n=1}^{N^{2}}\left\langle \left[ \varphi _{j+N^{2},m}\gamma
_{m,\downarrow }+\varphi _{j+N^{2},m+N^{2}}\gamma _{m,\uparrow }\right]
\right.   \notag \\
&&\times \left. \left[ \chi _{i,n}\gamma _{n,\downarrow }^{\dag }+\chi
_{i,n+N^{2}}\gamma _{n,\uparrow }^{\dag }\right] \right\rangle   \notag \\
&=&\sum_{m=1}^{2N^{2}}\varphi _{j+N^{2},m}\chi _{i,m}  \notag \\
&=&\left( \varphi \chi ^{T}\right) _{j+N^{2},i},
\end{eqnarray}%
where $\varphi $ and $\chi $ defined in Eq. (\ref{S}) are computed from the
exact diagonalization of the Hamiltonian matrix $\widetilde{H}$ in Eq. (\ref%
{H_tilde}).

\section*{Appendix B: Majorana lattice and edge modes}

\label{B} \setcounter{equation}{0} \renewcommand{\theequation}{B%
\arabic{equation}}

To gain intuition on the edge modes, we introduce the Majorana fermion
operators $a_{\mathbf{r},\sigma }=c_{\mathbf{r},\sigma }^{\dagger }+c_{%
\mathbf{r},\sigma }$, $b_{\mathbf{r},\sigma }=-i(c_{\mathbf{r},\sigma
}^{\dagger }-c_{\mathbf{r},\sigma })$, which satisfy the anticommutation
relations $\{a_{\mathbf{r},\sigma },a_{\mathbf{r}^{\prime },\sigma ^{\prime
}}\}=2\delta _{\mathbf{r},\mathbf{r}^{\prime }}\delta _{\sigma ,\sigma
^{\prime }}$, $\{b_{\mathbf{r},\sigma },b_{\mathbf{r}^{\prime },\sigma
^{\prime }}\}=2\delta _{\mathbf{r},\mathbf{r}^{\prime }}\delta _{\sigma
,\sigma ^{\prime }}$ and $\{a_{\mathbf{r},\sigma },b_{\mathbf{r}^{\prime
},\sigma ^{\prime }}\}=0$. The Majorana representation of the real space
Hamiltonian in Eq. (\ref{H_real_space}) is 
\begin{eqnarray}
H &=&\sum_{\mathbf{r}}\sum_{\mathbf{a}=\hat{x},\hat{y}}\Upsilon _{\mathbf{r}%
}^{\dagger }T_{\mathrm{NN}}\Upsilon _{\mathbf{r+a}}+\mathrm{H.c.}  \notag \\
&&+\sum_{\mathbf{r}}\Upsilon _{\mathbf{r}}^{\dagger }T_{\mathrm{O}}\Upsilon
_{\mathbf{r}},
\end{eqnarray}%
where $\Upsilon _{\mathbf{r}}^{\dagger }=\left( a_{\mathbf{r,\uparrow }}
\quad a_{\mathbf{r,\downarrow }} \quad b_{\mathbf{r,\uparrow }} \quad b_{%
\mathbf{r,\downarrow }}\right)$, and the coefficient matrix for nearest
neighbor and on-site pairing terms are 
\begin{equation}
T_{\mathrm{NN}}=\frac{1}{4}i\Delta _{-}\left( \sigma _{x}\otimes \sigma
_{+}\right) -\frac{1}{4}i\Delta _{+}\left( \sigma _{x}\otimes \sigma
_{-}\right) ,
\end{equation}
and 
\begin{equation}
T_{\mathrm{O}}=-\frac{1}{4}\Delta _{0}\left( \sigma _{x}\otimes \sigma
_{y}\right) ,
\end{equation}%
respectively, with $\sigma _{\pm }=\left( \sigma _{x}\pm \sigma _{y}\right)/2
$.

Next, we consider the $N\times N$ square lattice with cylindrical boundaries
condition: taking open boundary conditions in the $x$ direction and periodic
boundary condition in the $y$ direction. Then, take the following Fourier
transformations for the Majorana fermion operators:
\begin{equation}
\left( 
\begin{array}{c}
a_{m,k_{y},\sigma } \\ 
b_{m,k_{y},\sigma }%
\end{array}%
\right) =\frac{1}{\sqrt{N}}\sum_{n}e^{-ik_{y}n}\left( 
\begin{array}{c}
a_{m,n,\sigma } \\ 
b_{m,n,\sigma }%
\end{array}%
\right) ,
\end{equation}%
where $k_{y}=2\pi l/N,$ with $l$ $=0,1,...,N-1$. Note that in general, $%
a_{m,k_{y},\sigma }$ and $b_{m,k_{y},\sigma }$ are not Majorana fermion
operators, except at $k_{y}=0$ and $\pi $; we refer to such operators as
auxiliary operators. Then, the Hamiltonian can be written as $%
H=\sum_{k_{y}}H_{k_{y}}$, where 
\begin{eqnarray}
H_{k_{y}} &=&\sum_{m}\Upsilon _{m,k_{y}}^{\dagger }T_{\mathrm{NN}%
}^{k_{y}}\Upsilon _{m+1,k_{y}}+\mathrm{H.c.}  \notag \\
&&+\sum_{m}\Upsilon _{m,k_{y}}^{\dagger }T_{\mathrm{O}}^{k_{y}}\Upsilon
_{m,k_{y}},  \label{Hky}
\end{eqnarray}%
with $\Upsilon _{m,k_{y}}^{\dagger }=\left( a_{m,k_{y}\mathbf{,\uparrow }}
\quad a_{m,k_{y}\mathbf{,\downarrow }} \quad b_{m,k_{y}\mathbf{,\uparrow }}
\quad b_{m,k_{y}\mathbf{,\downarrow }}\right)$, and 
\begin{eqnarray}
T_{\mathrm{NN}}^{k_{y}} &=&T_{\mathrm{NN}},  \notag \\
T_{\mathrm{O}}^{k_{y}} &=&i\eta _{k_{y}}^{\ast }\left( \sigma _{x}\otimes
\sigma _{+}\right) -i\eta _{k_{y}}\left( \sigma _{x}\otimes \sigma
_{-}\right) ,  \notag \\
\eta _{k_{y}} &=&\frac{1}{4}(\Delta _{+}e^{-ik_{y}}+\Delta
_{-}e^{ik_{y}}+\Delta _{0}).
\end{eqnarray}%
We note that for each$\ k_{y}$, the Hamiltonian $H_{k_{y}}$ represents a
lattice of ladders about auxiliary operators $a_{m,k_{y},\sigma }$ and $%
b_{m,k_{y},\sigma }$. In Fig. \ref{kySpectrum}, we show the numerical
results of the single-particle spectra of $H_{k_{y}}$ for three sets of
typical parameters. We can see the existence of the flat-band zero modes
in the topological nontrivial case in Fig. \ref{kySpectrum} (a).

Actually, in the large $N$ limit, the Hamiltonian $H_{k_{y}}$ is expected to
possess zero energy edge modes, which can be determined by the following
matrix equation 
\begin{equation}
(T_{\mathrm{NN}}^{k_{y}})^{\dagger }\Psi _{m-1,k_{y}}+T_{\mathrm{O}%
}^{k_{y}}\Psi _{m,k_{y}}+T_{\mathrm{NN}}^{k_{y}}\Psi _{m+1,k_{y}}=0,
\end{equation}%
where $m=1,2,...N$, and $\Psi _{m,k_{y}}$ is a four-dimensional vector under
the basis of $\Upsilon _{m,k_{y}}$. The boundary condition is 
\begin{equation}
\Psi _{0,k_{y}}=0,\Psi _{N+1,k_{y}}=0.
\end{equation}

There are four zero energy edge modes when the winding number \cite%
{ryu2002topological, sun2012topological, wong2013majorana,
matsuura2013protected} of the pseudo spin Hamiltonians $H\left( \mathbf{k}%
\right) $ and $-H\left( -\mathbf{k}\right) $ in Eq. (\ref{H_BdG}) 
\begin{equation}
\mathcal{W}_{\pm }(k_{y})=\frac{1}{2\pi i}\int_{-\pi }^{\pi }dk_{x}\partial
_{k_{x}}\ln [\pm g(\pm \mathbf{k})],
\end{equation}%
are nonzero, where $g(\mathbf{k})= B_{x}(\mathbf{k})+iB_{y}(\mathbf{k}) $.
It can be checked that the condition for $\mathcal{W}_{\pm }(k_{y})$ to be
nonzero is $\left\vert p_{\pm }\right\vert <1$ and $\left\vert q_{\pm
}\right\vert <1$, where 
\begin{eqnarray}
p_{\pm } &=&-\frac{2\eta _{k_{y}}\pm \sqrt{4\eta _{k_{y}}^{2}-\Delta
_{+}\Delta _{-}}}{\Delta _{+}},  \notag \\
q_{\pm } &=&-\frac{2\eta _{k_{y}}^{\ast }\pm \sqrt{4(\eta _{k_{y}}^{\ast
})^{2}-\Delta _{+}\Delta _{-}}}{\Delta _{-}}.
\end{eqnarray}%
Under this condition, the zero modes are determined to be 
\begin{eqnarray}
\gamma _{k_{y}\mathbf{,\uparrow }} &=&A_{\mathbf{\uparrow }}\sum_{j=1}^{N}%
\left[ \left( p_{+}^{j}-p_{-}^{j}\right) a_{j,k_{y}\mathbf{,\uparrow }%
}\right.  \notag \\
&&\left. +i\left( p_{+}^{N-j+1}-p_{-}^{N-j+1}\right) b_{j,k_{y}\mathbf{%
,\uparrow }}\right] , \\
\gamma _{k_{y}\mathbf{,\downarrow }} &=&A_{\mathbf{\downarrow }%
}\sum_{j=1}^{N}\left[ \left( q_{+}^{j}-q_{-}^{j}\right) a_{j,k_{y}\mathbf{%
,\downarrow }}\right.  \notag \\
&&\left. +i\left( q_{+}^{N-j+1}-q_{-}^{N-j+1}\right) b_{j,k_{y}\mathbf{%
,\downarrow }}\right] ,
\end{eqnarray}%
and their corresponding Hermitian conjugate $\gamma _{k_{y}\mathbf{,}\sigma
}^{\dagger }$, in which $A_{\sigma }$ is a normalization constant. It can be
checked that the above edge zero mode operators are fermionic operators, since they
satisfy the anticommutation relations 
\begin{eqnarray}
\{\gamma _{k_{y}\mathbf{,}\sigma },\gamma _{k_{y}^{\prime }\mathbf{,}\sigma
^{\prime }}^{\dagger }\} &=&\delta _{k_{y},k_{y}^{\prime }}\delta _{\sigma
,\sigma ^{\prime }},  \notag \\
\{\gamma _{k_{y}\mathbf{,}\sigma },\gamma _{k_{y}^{\prime }\mathbf{,}\sigma
^{\prime }}\} &=&0.
\end{eqnarray}


\begin{thebibliography}{55}%
\makeatletter
\providecommand \@ifxundefined [1]{%
 \@ifx{#1\undefined}
}%
\providecommand \@ifnum [1]{%
 \ifnum #1\expandafter \@firstoftwo
 \else \expandafter \@secondoftwo
 \fi
}%
\providecommand \@ifx [1]{%
 \ifx #1\expandafter \@firstoftwo
 \else \expandafter \@secondoftwo
 \fi
}%
\providecommand \natexlab [1]{#1}%
\providecommand \enquote  [1]{``#1''}%
\providecommand \bibnamefont  [1]{#1}%
\providecommand \bibfnamefont [1]{#1}%
\providecommand \citenamefont [1]{#1}%
\providecommand \href@noop [0]{\@secondoftwo}%
\providecommand \href [0]{\begingroup \@sanitize@url \@href}%
\providecommand \@href[1]{\@@startlink{#1}\@@href}%
\providecommand \@@href[1]{\endgroup#1\@@endlink}%
\providecommand \@sanitize@url [0]{\catcode `\\12\catcode `\$12\catcode
  `\&12\catcode `\#12\catcode `\^12\catcode `\_12\catcode `\%12\relax}%
\providecommand \@@startlink[1]{}%
\providecommand \@@endlink[0]{}%
\providecommand \url  [0]{\begingroup\@sanitize@url \@url }%
\providecommand \@url [1]{\endgroup\@href {#1}{\urlprefix }}%
\providecommand \urlprefix  [0]{URL }%
\providecommand \Eprint [0]{\href }%
\providecommand \doibase [0]{https://doi.org/}%
\providecommand \selectlanguage [0]{\@gobble}%
\providecommand \bibinfo  [0]{\@secondoftwo}%
\providecommand \bibfield  [0]{\@secondoftwo}%
\providecommand \translation [1]{[#1]}%
\providecommand \BibitemOpen [0]{}%
\providecommand \bibitemStop [0]{}%
\providecommand \bibitemNoStop [0]{.\EOS\space}%
\providecommand \EOS [0]{\spacefactor3000\relax}%
\providecommand \BibitemShut  [1]{\csname bibitem#1\endcsname}%
\let\auto@bib@innerbib\@empty
\bibitem [{\citenamefont {Klitzing}\ \emph {et~al.}(1980)\citenamefont
  {Klitzing}, \citenamefont {Dorda},\ and\ \citenamefont
  {Pepper}}]{Klitzing1980}%
  \BibitemOpen
  \bibfield  {author} {\bibinfo {author} {\bibfnamefont {K.~v.}\ \bibnamefont
  {Klitzing}}, \bibinfo {author} {\bibfnamefont {G.}~\bibnamefont {Dorda}},\
  and\ \bibinfo {author} {\bibfnamefont {M.}~\bibnamefont {Pepper}},\
  }\bibfield  {title} {\bibinfo {title} {New method for high-accuracy
  determination of the fine-structure constant based on quantized hall
  resistance},\ }\href {https://doi.org/10.1103/PhysRevLett.45.494} {\bibfield
  {journal} {\bibinfo  {journal} {Phys. Rev. Lett.}\ }\textbf {\bibinfo
  {volume} {45}},\ \bibinfo {pages} {494} (\bibinfo {year} {1980})}\BibitemShut
  {NoStop}%
\bibitem [{\citenamefont {Laughlin}(1981)}]{Laughlin1981}%
  \BibitemOpen
  \bibfield  {author} {\bibinfo {author} {\bibfnamefont {R.~B.}\ \bibnamefont
  {Laughlin}},\ }\bibfield  {title} {\bibinfo {title} {Quantized hall
  conductivity in two dimensions},\ }\href
  {https://doi.org/10.1103/PhysRevB.23.5632} {\bibfield  {journal} {\bibinfo
  {journal} {Phys. Rev. B}\ }\textbf {\bibinfo {volume} {23}},\ \bibinfo
  {pages} {5632} (\bibinfo {year} {1981})}\BibitemShut {NoStop}%
\bibitem [{\citenamefont {Thouless}\ \emph {et~al.}(1982)\citenamefont
  {Thouless}, \citenamefont {Kohmoto}, \citenamefont {Nightingale},\ and\
  \citenamefont {den Nijs}}]{Thouless1982}%
  \BibitemOpen
  \bibfield  {author} {\bibinfo {author} {\bibfnamefont {D.~J.}\ \bibnamefont
  {Thouless}}, \bibinfo {author} {\bibfnamefont {M.}~\bibnamefont {Kohmoto}},
  \bibinfo {author} {\bibfnamefont {M.~P.}\ \bibnamefont {Nightingale}},\ and\
  \bibinfo {author} {\bibfnamefont {M.}~\bibnamefont {den Nijs}},\ }\bibfield
  {title} {\bibinfo {title} {Quantized hall conductance in a two-dimensional
  periodic potential},\ }\href {https://doi.org/10.1103/PhysRevLett.49.405}
  {\bibfield  {journal} {\bibinfo  {journal} {Phys. Rev. Lett.}\ }\textbf
  {\bibinfo {volume} {49}},\ \bibinfo {pages} {405} (\bibinfo {year}
  {1982})}\BibitemShut {NoStop}%
\bibitem [{\citenamefont {Laughlin}(1983)}]{Laughlin1983}%
  \BibitemOpen
  \bibfield  {author} {\bibinfo {author} {\bibfnamefont {R.~B.}\ \bibnamefont
  {Laughlin}},\ }\bibfield  {title} {\bibinfo {title} {Anomalous quantum hall
  effect: An incompressible quantum fluid with fractionally charged
  excitations},\ }\href {https://doi.org/10.1103/PhysRevLett.50.1395}
  {\bibfield  {journal} {\bibinfo  {journal} {Phys. Rev. Lett.}\ }\textbf
  {\bibinfo {volume} {50}},\ \bibinfo {pages} {1395} (\bibinfo {year}
  {1983})}\BibitemShut {NoStop}%
\bibitem [{\citenamefont {Haldane}(1988)}]{Haldane1988}%
  \BibitemOpen
  \bibfield  {author} {\bibinfo {author} {\bibfnamefont {F.~D.~M.}\
  \bibnamefont {Haldane}},\ }\bibfield  {title} {\bibinfo {title} {Model for a
  quantum hall effect without landau levels: Condensed-matter realization of
  the "parity anomaly"},\ }\href {https://doi.org/10.1103/PhysRevLett.61.2015}
  {\bibfield  {journal} {\bibinfo  {journal} {Phys. Rev. Lett.}\ }\textbf
  {\bibinfo {volume} {61}},\ \bibinfo {pages} {2015} (\bibinfo {year}
  {1988})}\BibitemShut {NoStop}%
\bibitem [{\citenamefont {Kane}\ and\ \citenamefont {Mele}(2005)}]{Kane2005}%
  \BibitemOpen
  \bibfield  {author} {\bibinfo {author} {\bibfnamefont {C.~L.}\ \bibnamefont
  {Kane}}\ and\ \bibinfo {author} {\bibfnamefont {E.~J.}\ \bibnamefont
  {Mele}},\ }\bibfield  {title} {\bibinfo {title} {Quantum spin hall effect in
  graphene},\ }\href {https://doi.org/10.1103/PhysRevLett.95.226801} {\bibfield
   {journal} {\bibinfo  {journal} {Phys. Rev. Lett.}\ }\textbf {\bibinfo
  {volume} {95}},\ \bibinfo {pages} {226801} (\bibinfo {year}
  {2005})}\BibitemShut {NoStop}%
\bibitem [{\citenamefont {Konig}\ \emph {et~al.}(2007)\citenamefont {Konig},
  \citenamefont {Wiedmann}, \citenamefont {Brune}, \citenamefont {Roth},
  \citenamefont {Buhmann}, \citenamefont {Molenkamp}, \citenamefont {Qi},\ and\
  \citenamefont {Zhang}}]{Konig2007}%
  \BibitemOpen
  \bibfield  {author} {\bibinfo {author} {\bibfnamefont {M.}~\bibnamefont
  {Konig}}, \bibinfo {author} {\bibfnamefont {S.}~\bibnamefont {Wiedmann}},
  \bibinfo {author} {\bibfnamefont {C.}~\bibnamefont {Brune}}, \bibinfo
  {author} {\bibfnamefont {A.}~\bibnamefont {Roth}}, \bibinfo {author}
  {\bibfnamefont {H.}~\bibnamefont {Buhmann}}, \bibinfo {author} {\bibfnamefont
  {L.~W.}\ \bibnamefont {Molenkamp}}, \bibinfo {author} {\bibfnamefont {X.-L.}\
  \bibnamefont {Qi}},\ and\ \bibinfo {author} {\bibfnamefont {S.-C.}\
  \bibnamefont {Zhang}},\ }\bibfield  {title} {\bibinfo {title} {Quantum spin
  hall insulator state in hgte quantum wells},\ }\href
  {https://www.science.org/doi/abs/10.1126/science.1148047} {\bibfield
  {journal} {\bibinfo  {journal} {Science}\ }\textbf {\bibinfo {volume}
  {318}},\ \bibinfo {pages} {766} (\bibinfo {year} {2007})}\BibitemShut
  {NoStop}%
\bibitem [{\citenamefont {Fu}\ and\ \citenamefont {Kane}(2007)}]{Fu2007a}%
  \BibitemOpen
  \bibfield  {author} {\bibinfo {author} {\bibfnamefont {L.}~\bibnamefont
  {Fu}}\ and\ \bibinfo {author} {\bibfnamefont {C.~L.}\ \bibnamefont {Kane}},\
  }\bibfield  {title} {\bibinfo {title} {Topological insulators with inversion
  symmetry},\ }\href {https://doi.org/10.1103/PhysRevB.76.045302} {\bibfield
  {journal} {\bibinfo  {journal} {Phys. Rev. B}\ }\textbf {\bibinfo {volume}
  {76}},\ \bibinfo {pages} {045302} (\bibinfo {year} {2007})}\BibitemShut
  {NoStop}%
\bibitem [{\citenamefont {Qi}\ \emph {et~al.}(2009)\citenamefont {Qi},
  \citenamefont {Hughes}, \citenamefont {Raghu},\ and\ \citenamefont
  {Zhang}}]{Qi2009}%
  \BibitemOpen
  \bibfield  {author} {\bibinfo {author} {\bibfnamefont {X.-L.}\ \bibnamefont
  {Qi}}, \bibinfo {author} {\bibfnamefont {T.~L.}\ \bibnamefont {Hughes}},
  \bibinfo {author} {\bibfnamefont {S.}~\bibnamefont {Raghu}},\ and\ \bibinfo
  {author} {\bibfnamefont {S.-C.}\ \bibnamefont {Zhang}},\ }\bibfield  {title}
  {\bibinfo {title} {Time-reversal-invariant topological superconductors and
  superfluids in two and three dimensions},\ }\href
  {https://doi.org/10.1103/PhysRevLett.102.187001} {\bibfield  {journal}
  {\bibinfo  {journal} {Phys. Rev. Lett.}\ }\textbf {\bibinfo {volume} {102}},\
  \bibinfo {pages} {187001} (\bibinfo {year} {2009})}\BibitemShut {NoStop}%
\bibitem [{\citenamefont {Qi}\ and\ \citenamefont
  {Zhang}(2011)}]{qi2011topological}%
  \BibitemOpen
  \bibfield  {author} {\bibinfo {author} {\bibfnamefont {X.-L.}\ \bibnamefont
  {Qi}}\ and\ \bibinfo {author} {\bibfnamefont {S.-C.}\ \bibnamefont {Zhang}},\
  }\bibfield  {title} {\bibinfo {title} {Topological insulators and
  superconductors},\ }\href {https://doi.org/10.1103/RevModPhys.83.1057}
  {\bibfield  {journal} {\bibinfo  {journal} {Rev. Mod. Phys.}\ }\textbf
  {\bibinfo {volume} {83}},\ \bibinfo {pages} {1057} (\bibinfo {year}
  {2011})}\BibitemShut {NoStop}%
\bibitem [{\citenamefont {Chang}\ \emph {et~al.}(2013)\citenamefont {Chang},
  \citenamefont {Zhang}, \citenamefont {Feng}, \citenamefont {Shen},
  \citenamefont {Zhang}, \citenamefont {Guo}, \citenamefont {Li}, \citenamefont
  {Ou}, \citenamefont {Wei}, \citenamefont {Wang} \emph {et~al.}}]{Chang2013}%
  \BibitemOpen
  \bibfield  {author} {\bibinfo {author} {\bibfnamefont {C.-Z.}\ \bibnamefont
  {Chang}}, \bibinfo {author} {\bibfnamefont {J.}~\bibnamefont {Zhang}},
  \bibinfo {author} {\bibfnamefont {X.}~\bibnamefont {Feng}}, \bibinfo {author}
  {\bibfnamefont {J.}~\bibnamefont {Shen}}, \bibinfo {author} {\bibfnamefont
  {Z.}~\bibnamefont {Zhang}}, \bibinfo {author} {\bibfnamefont
  {M.}~\bibnamefont {Guo}}, \bibinfo {author} {\bibfnamefont {K.}~\bibnamefont
  {Li}}, \bibinfo {author} {\bibfnamefont {Y.}~\bibnamefont {Ou}}, \bibinfo
  {author} {\bibfnamefont {P.}~\bibnamefont {Wei}}, \bibinfo {author}
  {\bibfnamefont {L.-L.}\ \bibnamefont {Wang}}, \emph {et~al.},\ }\bibfield
  {title} {\bibinfo {title} {Experimental observation of the quantum anomalous
  hall effect in a magnetic topological insulator},\ }\href
  {https://www.science.org/doi/abs/10.1126/science.1234414} {\bibfield
  {journal} {\bibinfo  {journal} {Science}\ }\textbf {\bibinfo {volume}
  {340}},\ \bibinfo {pages} {167} (\bibinfo {year} {2013})}\BibitemShut
  {NoStop}%
\bibitem [{\citenamefont {Wong}\ \emph {et~al.}(2013)\citenamefont {Wong},
  \citenamefont {Liu}, \citenamefont {Law},\ and\ \citenamefont
  {Lee}}]{wong2013majorana}%
  \BibitemOpen
  \bibfield  {author} {\bibinfo {author} {\bibfnamefont {C.~L.~M.}\
  \bibnamefont {Wong}}, \bibinfo {author} {\bibfnamefont {J.}~\bibnamefont
  {Liu}}, \bibinfo {author} {\bibfnamefont {K.~T.}\ \bibnamefont {Law}},\ and\
  \bibinfo {author} {\bibfnamefont {P.~A.}\ \bibnamefont {Lee}},\ }\bibfield
  {title} {\bibinfo {title} {Majorana flat bands and unidirectional majorana
  edge states in gapless topological superconductors},\ }\href
  {https://doi.org/10.1103/PhysRevB.88.060504} {\bibfield  {journal} {\bibinfo
  {journal} {Phys. Rev. B}\ }\textbf {\bibinfo {volume} {88}},\ \bibinfo
  {pages} {060504} (\bibinfo {year} {2013})}\BibitemShut {NoStop}%
\bibitem [{\citenamefont {Chiu}\ and\ \citenamefont
  {Schnyder}(2014)}]{Chiu2014}%
  \BibitemOpen
  \bibfield  {author} {\bibinfo {author} {\bibfnamefont {C.-K.}\ \bibnamefont
  {Chiu}}\ and\ \bibinfo {author} {\bibfnamefont {A.~P.}\ \bibnamefont
  {Schnyder}},\ }\bibfield  {title} {\bibinfo {title} {Classification of
  reflection-symmetry-protected topological semimetals and nodal
  superconductors},\ }\href {https://doi.org/10.1103/PhysRevB.90.205136}
  {\bibfield  {journal} {\bibinfo  {journal} {Phys. Rev. B}\ }\textbf {\bibinfo
  {volume} {90}},\ \bibinfo {pages} {205136} (\bibinfo {year}
  {2014})}\BibitemShut {NoStop}%
\bibitem [{\citenamefont {Liu}\ and\ \citenamefont
  {Wakabayashi}(2017)}]{Liu2017}%
  \BibitemOpen
  \bibfield  {author} {\bibinfo {author} {\bibfnamefont {F.}~\bibnamefont
  {Liu}}\ and\ \bibinfo {author} {\bibfnamefont {K.}~\bibnamefont
  {Wakabayashi}},\ }\bibfield  {title} {\bibinfo {title} {Novel topological
  phase with a zero berry curvature},\ }\href
  {https://doi.org/10.1103/PhysRevLett.118.076803} {\bibfield  {journal}
  {\bibinfo  {journal} {Phys. Rev. Lett.}\ }\textbf {\bibinfo {volume} {118}},\
  \bibinfo {pages} {076803} (\bibinfo {year} {2017})}\BibitemShut {NoStop}%
\bibitem [{\citenamefont {Zhang}\ \emph {et~al.}(2019)\citenamefont {Zhang},
  \citenamefont {Jiang}, \citenamefont {Song}, \citenamefont {Huang},
  \citenamefont {He}, \citenamefont {Fang}, \citenamefont {Weng},\ and\
  \citenamefont {Fang}}]{Zhang2019}%
  \BibitemOpen
  \bibfield  {author} {\bibinfo {author} {\bibfnamefont {T.}~\bibnamefont
  {Zhang}}, \bibinfo {author} {\bibfnamefont {Y.}~\bibnamefont {Jiang}},
  \bibinfo {author} {\bibfnamefont {Z.}~\bibnamefont {Song}}, \bibinfo {author}
  {\bibfnamefont {H.}~\bibnamefont {Huang}}, \bibinfo {author} {\bibfnamefont
  {Y.}~\bibnamefont {He}}, \bibinfo {author} {\bibfnamefont {Z.}~\bibnamefont
  {Fang}}, \bibinfo {author} {\bibfnamefont {H.}~\bibnamefont {Weng}},\ and\
  \bibinfo {author} {\bibfnamefont {C.}~\bibnamefont {Fang}},\ }\bibfield
  {title} {\bibinfo {title} {Catalogue of topological electronic materials},\
  }\href {https://www.nature.com/articles/s41586-019-0944-6} {\bibfield
  {journal} {\bibinfo  {journal} {Nature}\ }\textbf {\bibinfo {volume} {566}},\
  \bibinfo {pages} {475} (\bibinfo {year} {2019})}\BibitemShut {NoStop}%
\bibitem [{\citenamefont {Xie}\ \emph {et~al.}(2023)\citenamefont {Xie},
  \citenamefont {Jin},\ and\ \citenamefont {Song}}]{xie2023antihelical}%
  \BibitemOpen
  \bibfield  {author} {\bibinfo {author} {\bibfnamefont {L.}~\bibnamefont
  {Xie}}, \bibinfo {author} {\bibfnamefont {L.}~\bibnamefont {Jin}},\ and\
  \bibinfo {author} {\bibfnamefont {Z.}~\bibnamefont {Song}},\ }\bibfield
  {title} {\bibinfo {title} {Antihelical edge states in two-dimensional
  photonic topological metals},\ }\href
  {https://doi.org/https://doi.org/10.1016/j.scib.2023.01.018} {\bibfield
  {journal} {\bibinfo  {journal} {Sci. Bull.}\ }\textbf {\bibinfo {volume}
  {68}},\ \bibinfo {pages} {255} (\bibinfo {year} {2023})}\BibitemShut
  {NoStop}%
\bibitem [{\citenamefont {Chiu}\ \emph {et~al.}(2016)\citenamefont {Chiu},
  \citenamefont {Teo}, \citenamefont {Schnyder},\ and\ \citenamefont
  {Ryu}}]{chiu2016classification}%
  \BibitemOpen
  \bibfield  {author} {\bibinfo {author} {\bibfnamefont {C.-K.}\ \bibnamefont
  {Chiu}}, \bibinfo {author} {\bibfnamefont {J.~C.~Y.}\ \bibnamefont {Teo}},
  \bibinfo {author} {\bibfnamefont {A.~P.}\ \bibnamefont {Schnyder}},\ and\
  \bibinfo {author} {\bibfnamefont {S.}~\bibnamefont {Ryu}},\ }\bibfield
  {title} {\bibinfo {title} {Classification of topological quantum matter with
  symmetries},\ }\href {https://doi.org/10.1103/RevModPhys.88.035005}
  {\bibfield  {journal} {\bibinfo  {journal} {Rev. Mod. Phys.}\ }\textbf
  {\bibinfo {volume} {88}},\ \bibinfo {pages} {035005} (\bibinfo {year}
  {2016})}\BibitemShut {NoStop}%
\bibitem [{\citenamefont {Schnyder}\ \emph {et~al.}(2008)\citenamefont
  {Schnyder}, \citenamefont {Ryu}, \citenamefont {Furusaki},\ and\
  \citenamefont {Ludwig}}]{Schnyder2008}%
  \BibitemOpen
  \bibfield  {author} {\bibinfo {author} {\bibfnamefont {A.~P.}\ \bibnamefont
  {Schnyder}}, \bibinfo {author} {\bibfnamefont {S.}~\bibnamefont {Ryu}},
  \bibinfo {author} {\bibfnamefont {A.}~\bibnamefont {Furusaki}},\ and\
  \bibinfo {author} {\bibfnamefont {A.~W.~W.}\ \bibnamefont {Ludwig}},\
  }\bibfield  {title} {\bibinfo {title} {Classification of topological
  insulators and superconductors in three spatial dimensions},\ }\href
  {https://doi.org/10.1103/PhysRevB.78.195125} {\bibfield  {journal} {\bibinfo
  {journal} {Phys. Rev. B}\ }\textbf {\bibinfo {volume} {78}},\ \bibinfo
  {pages} {195125} (\bibinfo {year} {2008})}\BibitemShut {NoStop}%
\bibitem [{\citenamefont {B\'eri}(2010)}]{beri2010topologically}%
  \BibitemOpen
  \bibfield  {author} {\bibinfo {author} {\bibfnamefont {B.}~\bibnamefont
  {B\'eri}},\ }\bibfield  {title} {\bibinfo {title} {Topologically stable
  gapless phases of time-reversal-invariant superconductors},\ }\href
  {https://doi.org/10.1103/PhysRevB.81.134515} {\bibfield  {journal} {\bibinfo
  {journal} {Phys. Rev. B}\ }\textbf {\bibinfo {volume} {81}},\ \bibinfo
  {pages} {134515} (\bibinfo {year} {2010})}\BibitemShut {NoStop}%
\bibitem [{\citenamefont {Queiroz}\ and\ \citenamefont
  {Schnyder}(2014)}]{queiroz2014stability}%
  \BibitemOpen
  \bibfield  {author} {\bibinfo {author} {\bibfnamefont {R.}~\bibnamefont
  {Queiroz}}\ and\ \bibinfo {author} {\bibfnamefont {A.~P.}\ \bibnamefont
  {Schnyder}},\ }\bibfield  {title} {\bibinfo {title} {Stability of flat-band
  edge states in topological superconductors without inversion center},\ }\href
  {https://doi.org/10.1103/PhysRevB.89.054501} {\bibfield  {journal} {\bibinfo
  {journal} {Phys. Rev. B}\ }\textbf {\bibinfo {volume} {89}},\ \bibinfo
  {pages} {054501} (\bibinfo {year} {2014})}\BibitemShut {NoStop}%
\bibitem [{\citenamefont {Schnyder}\ and\ \citenamefont
  {Brydon}(2015)}]{schnyder2015topological}%
  \BibitemOpen
  \bibfield  {author} {\bibinfo {author} {\bibfnamefont {A.~P.}\ \bibnamefont
  {Schnyder}}\ and\ \bibinfo {author} {\bibfnamefont {P.~M.}\ \bibnamefont
  {Brydon}},\ }\bibfield  {title} {\bibinfo {title} {Topological surface states
  in nodal superconductors},\ }\href
  {https://iopscience.iop.org/article/10.1088/0953-8984/27/24/243201/meta}
  {\bibfield  {journal} {\bibinfo  {journal} {Journal of Physics: Condensed
  Matter}\ }\textbf {\bibinfo {volume} {27}},\ \bibinfo {pages} {243201}
  (\bibinfo {year} {2015})}\BibitemShut {NoStop}%
\bibitem [{\citenamefont {Bouhon}\ \emph {et~al.}(2018)\citenamefont {Bouhon},
  \citenamefont {Schmidt},\ and\ \citenamefont
  {Black-Schaffer}}]{bouhon2018topological}%
  \BibitemOpen
  \bibfield  {author} {\bibinfo {author} {\bibfnamefont {A.}~\bibnamefont
  {Bouhon}}, \bibinfo {author} {\bibfnamefont {J.}~\bibnamefont {Schmidt}},\
  and\ \bibinfo {author} {\bibfnamefont {A.~M.}\ \bibnamefont
  {Black-Schaffer}},\ }\bibfield  {title} {\bibinfo {title} {Topological nodal
  superconducting phases and topological phase transition in the hyperhoneycomb
  lattice},\ }\href {https://doi.org/10.1103/PhysRevB.97.104508} {\bibfield
  {journal} {\bibinfo  {journal} {Phys. Rev. B}\ }\textbf {\bibinfo {volume}
  {97}},\ \bibinfo {pages} {104508} (\bibinfo {year} {2018})}\BibitemShut
  {NoStop}%
\bibitem [{\citenamefont {Kobayashi}\ \emph {et~al.}(2018)\citenamefont
  {Kobayashi}, \citenamefont {Sumita}, \citenamefont {Yanase},\ and\
  \citenamefont {Sato}}]{kobayashi2018symmetry}%
  \BibitemOpen
  \bibfield  {author} {\bibinfo {author} {\bibfnamefont {S.}~\bibnamefont
  {Kobayashi}}, \bibinfo {author} {\bibfnamefont {S.}~\bibnamefont {Sumita}},
  \bibinfo {author} {\bibfnamefont {Y.}~\bibnamefont {Yanase}},\ and\ \bibinfo
  {author} {\bibfnamefont {M.}~\bibnamefont {Sato}},\ }\bibfield  {title}
  {\bibinfo {title} {Symmetry-protected line nodes and majorana flat bands in
  nodal crystalline superconductors},\ }\href
  {https://doi.org/10.1103/PhysRevB.97.180504} {\bibfield  {journal} {\bibinfo
  {journal} {Phys. Rev. B}\ }\textbf {\bibinfo {volume} {97}},\ \bibinfo
  {pages} {180504} (\bibinfo {year} {2018})}\BibitemShut {NoStop}%
\bibitem [{\citenamefont {Nayak}\ \emph {et~al.}(2021)\citenamefont {Nayak},
  \citenamefont {Steinbok}, \citenamefont {Roet}, \citenamefont {Koo},
  \citenamefont {Margalit}, \citenamefont {Feldman}, \citenamefont {Almoalem},
  \citenamefont {Kanigel}, \citenamefont {Fiete}, \citenamefont {Yan} \emph
  {et~al.}}]{nayak2021evidence}%
  \BibitemOpen
  \bibfield  {author} {\bibinfo {author} {\bibfnamefont {A.~K.}\ \bibnamefont
  {Nayak}}, \bibinfo {author} {\bibfnamefont {A.}~\bibnamefont {Steinbok}},
  \bibinfo {author} {\bibfnamefont {Y.}~\bibnamefont {Roet}}, \bibinfo {author}
  {\bibfnamefont {J.}~\bibnamefont {Koo}}, \bibinfo {author} {\bibfnamefont
  {G.}~\bibnamefont {Margalit}}, \bibinfo {author} {\bibfnamefont
  {I.}~\bibnamefont {Feldman}}, \bibinfo {author} {\bibfnamefont
  {A.}~\bibnamefont {Almoalem}}, \bibinfo {author} {\bibfnamefont
  {A.}~\bibnamefont {Kanigel}}, \bibinfo {author} {\bibfnamefont {G.~A.}\
  \bibnamefont {Fiete}}, \bibinfo {author} {\bibfnamefont {B.}~\bibnamefont
  {Yan}}, \emph {et~al.},\ }\bibfield  {title} {\bibinfo {title} {Evidence of
  topological boundary modes with topological nodal-point superconductivity},\
  }\href {https://www.nature.com/articles/s41567-021-01376-z} {\bibfield
  {journal} {\bibinfo  {journal} {Nature physics}\ }\textbf {\bibinfo {volume}
  {17}},\ \bibinfo {pages} {1413} (\bibinfo {year} {2021})}\BibitemShut
  {NoStop}%
\bibitem [{\citenamefont {Xie}\ \emph {et~al.}(2021)\citenamefont {Xie},
  \citenamefont {Wu}, \citenamefont {Jin},\ and\ \citenamefont
  {Song}}]{xie2021time}%
  \BibitemOpen
  \bibfield  {author} {\bibinfo {author} {\bibfnamefont {L.~C.}\ \bibnamefont
  {Xie}}, \bibinfo {author} {\bibfnamefont {H.~C.}\ \bibnamefont {Wu}},
  \bibinfo {author} {\bibfnamefont {L.}~\bibnamefont {Jin}},\ and\ \bibinfo
  {author} {\bibfnamefont {Z.}~\bibnamefont {Song}},\ }\bibfield  {title}
  {\bibinfo {title} {Time-reversal symmetric topological metal},\ }\href
  {https://doi.org/10.1103/PhysRevB.104.165422} {\bibfield  {journal} {\bibinfo
   {journal} {Phys. Rev. B}\ }\textbf {\bibinfo {volume} {104}},\ \bibinfo
  {pages} {165422} (\bibinfo {year} {2021})}\BibitemShut {NoStop}%
\bibitem [{\citenamefont {Bazarnik}\ \emph {et~al.}(2023)\citenamefont
  {Bazarnik}, \citenamefont {Lo~Conte}, \citenamefont {Mascot}, \citenamefont
  {von Bergmann}, \citenamefont {Morr},\ and\ \citenamefont
  {Wiesendanger}}]{bazarnik2023antiferromagnetism}%
  \BibitemOpen
  \bibfield  {author} {\bibinfo {author} {\bibfnamefont {M.}~\bibnamefont
  {Bazarnik}}, \bibinfo {author} {\bibfnamefont {R.}~\bibnamefont {Lo~Conte}},
  \bibinfo {author} {\bibfnamefont {E.}~\bibnamefont {Mascot}}, \bibinfo
  {author} {\bibfnamefont {K.}~\bibnamefont {von Bergmann}}, \bibinfo {author}
  {\bibfnamefont {D.~K.}\ \bibnamefont {Morr}},\ and\ \bibinfo {author}
  {\bibfnamefont {R.}~\bibnamefont {Wiesendanger}},\ }\bibfield  {title}
  {\bibinfo {title} {Antiferromagnetism-driven two-dimensional topological
  nodal-point superconductivity},\ }\href
  {https://www.nature.com/articles/s41467-023-36201-z} {\bibfield  {journal}
  {\bibinfo  {journal} {Nature Communications}\ }\textbf {\bibinfo {volume}
  {14}},\ \bibinfo {pages} {614} (\bibinfo {year} {2023})}\BibitemShut
  {NoStop}%
\bibitem [{\citenamefont {Mineev}\ and\ \citenamefont
  {Samokhin}(1999)}]{mineev1999introduction}%
  \BibitemOpen
  \bibfield  {author} {\bibinfo {author} {\bibfnamefont {V.~P.}\ \bibnamefont
  {Mineev}}\ and\ \bibinfo {author} {\bibfnamefont {K.~V.}\ \bibnamefont
  {Samokhin}},\ }\href@noop {} {\emph {\bibinfo {title} {Introduction to
  unconventional superconductivity}}}\ (\bibinfo  {publisher} {Gordon and
  Breach, New York},\ \bibinfo {year} {1999})\BibitemShut {NoStop}%
\bibitem [{\citenamefont {Gor'kov}\ and\ \citenamefont
  {Rashba}(2001)}]{gor2001superconducting}%
  \BibitemOpen
  \bibfield  {author} {\bibinfo {author} {\bibfnamefont {L.~P.}\ \bibnamefont
  {Gor'kov}}\ and\ \bibinfo {author} {\bibfnamefont {E.~I.}\ \bibnamefont
  {Rashba}},\ }\bibfield  {title} {\bibinfo {title} {Superconducting 2d system
  with lifted spin degeneracy: Mixed singlet-triplet state},\ }\href
  {https://doi.org/10.1103/PhysRevLett.87.037004} {\bibfield  {journal}
  {\bibinfo  {journal} {Phys. Rev. Lett.}\ }\textbf {\bibinfo {volume} {87}},\
  \bibinfo {pages} {037004} (\bibinfo {year} {2001})}\BibitemShut {NoStop}%
\bibitem [{\citenamefont {Aperis}\ \emph {et~al.}(2008)\citenamefont {Aperis},
  \citenamefont {Varelogiannis}, \citenamefont {Littlewood},\ and\
  \citenamefont {Simons}}]{aperis2008coexistence}%
  \BibitemOpen
  \bibfield  {author} {\bibinfo {author} {\bibfnamefont {A.}~\bibnamefont
  {Aperis}}, \bibinfo {author} {\bibfnamefont {G.}~\bibnamefont
  {Varelogiannis}}, \bibinfo {author} {\bibfnamefont {P.}~\bibnamefont
  {Littlewood}},\ and\ \bibinfo {author} {\bibfnamefont {B.}~\bibnamefont
  {Simons}},\ }\bibfield  {title} {\bibinfo {title} {Coexistence of spin
  density wave, d-wave singlet and staggered $\pi$-triplet superconductivity},\
  }\href
  {https://iopscience.iop.org/article/10.1088/0953-8984/20/43/434235/meta}
  {\bibfield  {journal} {\bibinfo  {journal} {Journal of Physics: Condensed
  Matter}\ }\textbf {\bibinfo {volume} {20}},\ \bibinfo {pages} {434235}
  (\bibinfo {year} {2008})}\BibitemShut {NoStop}%
\bibitem [{\citenamefont {Bergeret}\ and\ \citenamefont
  {Tokatly}(2013)}]{bergeret2013singlet}%
  \BibitemOpen
  \bibfield  {author} {\bibinfo {author} {\bibfnamefont {F.~S.}\ \bibnamefont
  {Bergeret}}\ and\ \bibinfo {author} {\bibfnamefont {I.~V.}\ \bibnamefont
  {Tokatly}},\ }\bibfield  {title} {\bibinfo {title} {Singlet-triplet
  conversion and the long-range proximity effect in superconductor-ferromagnet
  structures with generic spin dependent fields},\ }\href
  {https://doi.org/10.1103/PhysRevLett.110.117003} {\bibfield  {journal}
  {\bibinfo  {journal} {Phys. Rev. Lett.}\ }\textbf {\bibinfo {volume} {110}},\
  \bibinfo {pages} {117003} (\bibinfo {year} {2013})}\BibitemShut {NoStop}%
\bibitem [{\citenamefont {Wang}\ \emph {et~al.}(2022)\citenamefont {Wang},
  \citenamefont {Dvir}, \citenamefont {Mazur}, \citenamefont {Liu},
  \citenamefont {van Loo}, \citenamefont {Ten~Haaf}, \citenamefont {Bordin},
  \citenamefont {Gazibegovic}, \citenamefont {Badawy}, \citenamefont {Bakkers}
  \emph {et~al.}}]{wang2022singlet}%
  \BibitemOpen
  \bibfield  {author} {\bibinfo {author} {\bibfnamefont {G.}~\bibnamefont
  {Wang}}, \bibinfo {author} {\bibfnamefont {T.}~\bibnamefont {Dvir}}, \bibinfo
  {author} {\bibfnamefont {G.~P.}\ \bibnamefont {Mazur}}, \bibinfo {author}
  {\bibfnamefont {C.-X.}\ \bibnamefont {Liu}}, \bibinfo {author} {\bibfnamefont
  {N.}~\bibnamefont {van Loo}}, \bibinfo {author} {\bibfnamefont {S.~L.}\
  \bibnamefont {Ten~Haaf}}, \bibinfo {author} {\bibfnamefont {A.}~\bibnamefont
  {Bordin}}, \bibinfo {author} {\bibfnamefont {S.}~\bibnamefont {Gazibegovic}},
  \bibinfo {author} {\bibfnamefont {G.}~\bibnamefont {Badawy}}, \bibinfo
  {author} {\bibfnamefont {E.~P.}\ \bibnamefont {Bakkers}}, \emph {et~al.},\
  }\bibfield  {title} {\bibinfo {title} {Singlet and triplet cooper pair
  splitting in hybrid superconducting nanowires},\ }\href
  {https://www.nature.com/articles/s41586-022-05352-2} {\bibfield  {journal}
  {\bibinfo  {journal} {Nature}\ }\textbf {\bibinfo {volume} {612}},\ \bibinfo
  {pages} {448} (\bibinfo {year} {2022})}\BibitemShut {NoStop}%
\bibitem [{\citenamefont {Hatsugai}(1993)}]{hatsugai1993chern}%
  \BibitemOpen
  \bibfield  {author} {\bibinfo {author} {\bibfnamefont {Y.}~\bibnamefont
  {Hatsugai}},\ }\bibfield  {title} {\bibinfo {title} {Chern number and edge
  states in the integer quantum hall effect},\ }\href
  {https://doi.org/10.1103/PhysRevLett.71.3697} {\bibfield  {journal} {\bibinfo
   {journal} {Phys. Rev. Lett.}\ }\textbf {\bibinfo {volume} {71}},\ \bibinfo
  {pages} {3697} (\bibinfo {year} {1993})}\BibitemShut {NoStop}%
\bibitem [{\citenamefont {Kellendonk}\ \emph {et~al.}(2002)\citenamefont
  {Kellendonk}, \citenamefont {Richter},\ and\ \citenamefont
  {Schulz-Baldes}}]{kellendonk2002edge}%
  \BibitemOpen
  \bibfield  {author} {\bibinfo {author} {\bibfnamefont {J.}~\bibnamefont
  {Kellendonk}}, \bibinfo {author} {\bibfnamefont {T.}~\bibnamefont
  {Richter}},\ and\ \bibinfo {author} {\bibfnamefont {H.}~\bibnamefont
  {Schulz-Baldes}},\ }\bibfield  {title} {\bibinfo {title} {Edge current
  channels and chern numbers in the integer quantum hall effect},\ }\href
  {https://www.worldscientific.com/doi/abs/10.1142/S0129055X02001107}
  {\bibfield  {journal} {\bibinfo  {journal} {Reviews in Mathematical Physics}\
  }\textbf {\bibinfo {volume} {14}},\ \bibinfo {pages} {87} (\bibinfo {year}
  {2002})}\BibitemShut {NoStop}%
\bibitem [{\citenamefont {Qi}\ \emph {et~al.}(2006)\citenamefont {Qi},
  \citenamefont {Wu},\ and\ \citenamefont {Zhang}}]{qi2006general}%
  \BibitemOpen
  \bibfield  {author} {\bibinfo {author} {\bibfnamefont {X.-L.}\ \bibnamefont
  {Qi}}, \bibinfo {author} {\bibfnamefont {Y.-S.}\ \bibnamefont {Wu}},\ and\
  \bibinfo {author} {\bibfnamefont {S.-C.}\ \bibnamefont {Zhang}},\ }\bibfield
  {title} {\bibinfo {title} {General theorem relating the bulk topological
  number to edge states in two-dimensional insulators},\ }\href
  {https://doi.org/10.1103/PhysRevB.74.045125} {\bibfield  {journal} {\bibinfo
  {journal} {Phys. Rev. B}\ }\textbf {\bibinfo {volume} {74}},\ \bibinfo
  {pages} {045125} (\bibinfo {year} {2006})}\BibitemShut {NoStop}%
\bibitem [{\citenamefont {Mong}\ and\ \citenamefont
  {Shivamoggi}(2011)}]{mong2011edge}%
  \BibitemOpen
  \bibfield  {author} {\bibinfo {author} {\bibfnamefont {R.~S.~K.}\
  \bibnamefont {Mong}}\ and\ \bibinfo {author} {\bibfnamefont {V.}~\bibnamefont
  {Shivamoggi}},\ }\bibfield  {title} {\bibinfo {title} {Edge states and the
  bulk-boundary correspondence in dirac hamiltonians},\ }\href
  {https://doi.org/10.1103/PhysRevB.83.125109} {\bibfield  {journal} {\bibinfo
  {journal} {Phys. Rev. B}\ }\textbf {\bibinfo {volume} {83}},\ \bibinfo
  {pages} {125109} (\bibinfo {year} {2011})}\BibitemShut {NoStop}%
\bibitem [{\citenamefont {Essin}\ and\ \citenamefont
  {Gurarie}(2011)}]{essin2011bulk}%
  \BibitemOpen
  \bibfield  {author} {\bibinfo {author} {\bibfnamefont {A.~M.}\ \bibnamefont
  {Essin}}\ and\ \bibinfo {author} {\bibfnamefont {V.}~\bibnamefont
  {Gurarie}},\ }\bibfield  {title} {\bibinfo {title} {Bulk-boundary
  correspondence of topological insulators from their respective green's
  functions},\ }\href {https://doi.org/10.1103/PhysRevB.84.125132} {\bibfield
  {journal} {\bibinfo  {journal} {Phys. Rev. B}\ }\textbf {\bibinfo {volume}
  {84}},\ \bibinfo {pages} {125132} (\bibinfo {year} {2011})}\BibitemShut
  {NoStop}%
\bibitem [{\citenamefont {Bianco}\ and\ \citenamefont
  {Resta}(2011)}]{bianco2011mapping}%
  \BibitemOpen
  \bibfield  {author} {\bibinfo {author} {\bibfnamefont {R.}~\bibnamefont
  {Bianco}}\ and\ \bibinfo {author} {\bibfnamefont {R.}~\bibnamefont {Resta}},\
  }\bibfield  {title} {\bibinfo {title} {Mapping topological order in
  coordinate space},\ }\href {https://doi.org/10.1103/PhysRevB.84.241106}
  {\bibfield  {journal} {\bibinfo  {journal} {Phys. Rev. B}\ }\textbf {\bibinfo
  {volume} {84}},\ \bibinfo {pages} {241106} (\bibinfo {year}
  {2011})}\BibitemShut {NoStop}%
\bibitem [{\citenamefont {Sykes}\ and\ \citenamefont
  {Barnett}(2021)}]{sykes2021local}%
  \BibitemOpen
  \bibfield  {author} {\bibinfo {author} {\bibfnamefont {J.}~\bibnamefont
  {Sykes}}\ and\ \bibinfo {author} {\bibfnamefont {R.}~\bibnamefont
  {Barnett}},\ }\bibfield  {title} {\bibinfo {title} {Local topological markers
  in odd dimensions},\ }\href {https://doi.org/10.1103/PhysRevB.103.155134}
  {\bibfield  {journal} {\bibinfo  {journal} {Phys. Rev. B}\ }\textbf {\bibinfo
  {volume} {103}},\ \bibinfo {pages} {155134} (\bibinfo {year}
  {2021})}\BibitemShut {NoStop}%
\bibitem [{\citenamefont {Chen}(2023{\natexlab{a}})}]{chen2023optical}%
  \BibitemOpen
  \bibfield  {author} {\bibinfo {author} {\bibfnamefont {W.}~\bibnamefont
  {Chen}},\ }\bibfield  {title} {\bibinfo {title} {Optical absorption
  measurement of spin berry curvature and spin chern marker},\ }\href
  {https://iopscience.iop.org/article/10.1088/1361-648X/acba72/meta} {\bibfield
   {journal} {\bibinfo  {journal} {Journal of Physics: Condensed Matter}\
  }\textbf {\bibinfo {volume} {35}},\ \bibinfo {pages} {155601} (\bibinfo
  {year} {2023}{\natexlab{a}})}\BibitemShut {NoStop}%
\bibitem [{\citenamefont {Chen}(2023{\natexlab{b}})}]{chen2023universal}%
  \BibitemOpen
  \bibfield  {author} {\bibinfo {author} {\bibfnamefont {W.}~\bibnamefont
  {Chen}},\ }\bibfield  {title} {\bibinfo {title} {Universal topological
  marker},\ }\href {https://doi.org/10.1103/PhysRevB.107.045111} {\bibfield
  {journal} {\bibinfo  {journal} {Phys. Rev. B}\ }\textbf {\bibinfo {volume}
  {107}},\ \bibinfo {pages} {045111} (\bibinfo {year}
  {2023}{\natexlab{b}})}\BibitemShut {NoStop}%
\bibitem [{\citenamefont {Ringel}\ and\ \citenamefont
  {Kraus}(2011)}]{ringel2011determining}%
  \BibitemOpen
  \bibfield  {author} {\bibinfo {author} {\bibfnamefont {Z.}~\bibnamefont
  {Ringel}}\ and\ \bibinfo {author} {\bibfnamefont {Y.~E.}\ \bibnamefont
  {Kraus}},\ }\bibfield  {title} {\bibinfo {title} {Determining topological
  order from a local ground-state correlation function},\ }\href
  {https://doi.org/10.1103/PhysRevB.83.245115} {\bibfield  {journal} {\bibinfo
  {journal} {Phys. Rev. B}\ }\textbf {\bibinfo {volume} {83}},\ \bibinfo
  {pages} {245115} (\bibinfo {year} {2011})}\BibitemShut {NoStop}%
\bibitem [{\citenamefont {Lepori}\ \emph {et~al.}(2023)\citenamefont {Lepori},
  \citenamefont {Burrello}, \citenamefont {Trombettoni},\ and\ \citenamefont
  {Paganelli}}]{lepori2023strange}%
  \BibitemOpen
  \bibfield  {author} {\bibinfo {author} {\bibfnamefont {L.}~\bibnamefont
  {Lepori}}, \bibinfo {author} {\bibfnamefont {M.}~\bibnamefont {Burrello}},
  \bibinfo {author} {\bibfnamefont {A.}~\bibnamefont {Trombettoni}},\ and\
  \bibinfo {author} {\bibfnamefont {S.}~\bibnamefont {Paganelli}},\ }\bibfield
  {title} {\bibinfo {title} {Strange correlators for topological quantum
  systems from bulk-boundary correspondence},\ }\href
  {https://doi.org/10.1103/PhysRevB.108.035110} {\bibfield  {journal} {\bibinfo
   {journal} {Phys. Rev. B}\ }\textbf {\bibinfo {volume} {108}},\ \bibinfo
  {pages} {035110} (\bibinfo {year} {2023})}\BibitemShut {NoStop}%
\bibitem [{\citenamefont {Li}\ and\ \citenamefont
  {Haldane}(2008)}]{li2008entanglement}%
  \BibitemOpen
  \bibfield  {author} {\bibinfo {author} {\bibfnamefont {H.}~\bibnamefont
  {Li}}\ and\ \bibinfo {author} {\bibfnamefont {F.~D.~M.}\ \bibnamefont
  {Haldane}},\ }\bibfield  {title} {\bibinfo {title} {Entanglement spectrum as
  a generalization of entanglement entropy: Identification of topological order
  in non-abelian fractional quantum hall effect states},\ }\href
  {https://doi.org/10.1103/PhysRevLett.101.010504} {\bibfield  {journal}
  {\bibinfo  {journal} {Phys. Rev. Lett.}\ }\textbf {\bibinfo {volume} {101}},\
  \bibinfo {pages} {010504} (\bibinfo {year} {2008})}\BibitemShut {NoStop}%
\bibitem [{\citenamefont {Turner}\ \emph {et~al.}(2010)\citenamefont {Turner},
  \citenamefont {Zhang},\ and\ \citenamefont
  {Vishwanath}}]{turner2010entanglement}%
  \BibitemOpen
  \bibfield  {author} {\bibinfo {author} {\bibfnamefont {A.~M.}\ \bibnamefont
  {Turner}}, \bibinfo {author} {\bibfnamefont {Y.}~\bibnamefont {Zhang}},\ and\
  \bibinfo {author} {\bibfnamefont {A.}~\bibnamefont {Vishwanath}},\ }\bibfield
   {title} {\bibinfo {title} {Entanglement and inversion symmetry in
  topological insulators},\ }\href {https://doi.org/10.1103/PhysRevB.82.241102}
  {\bibfield  {journal} {\bibinfo  {journal} {Phys. Rev. B}\ }\textbf {\bibinfo
  {volume} {82}},\ \bibinfo {pages} {241102} (\bibinfo {year}
  {2010})}\BibitemShut {NoStop}%
\bibitem [{\citenamefont {Prodan}\ \emph {et~al.}(2010)\citenamefont {Prodan},
  \citenamefont {Hughes},\ and\ \citenamefont
  {Bernevig}}]{prodan2010entanglement}%
  \BibitemOpen
  \bibfield  {author} {\bibinfo {author} {\bibfnamefont {E.}~\bibnamefont
  {Prodan}}, \bibinfo {author} {\bibfnamefont {T.~L.}\ \bibnamefont {Hughes}},\
  and\ \bibinfo {author} {\bibfnamefont {B.~A.}\ \bibnamefont {Bernevig}},\
  }\bibfield  {title} {\bibinfo {title} {Entanglement spectrum of a disordered
  topological chern insulator},\ }\href
  {https://doi.org/10.1103/PhysRevLett.105.115501} {\bibfield  {journal}
  {\bibinfo  {journal} {Phys. Rev. Lett.}\ }\textbf {\bibinfo {volume} {105}},\
  \bibinfo {pages} {115501} (\bibinfo {year} {2010})}\BibitemShut {NoStop}%
\bibitem [{\citenamefont {Li}\ \emph {et~al.}(2009)\citenamefont {Li},
  \citenamefont {Chu}, \citenamefont {Jain},\ and\ \citenamefont
  {Shen}}]{Li2009}%
  \BibitemOpen
  \bibfield  {author} {\bibinfo {author} {\bibfnamefont {J.}~\bibnamefont
  {Li}}, \bibinfo {author} {\bibfnamefont {R.-L.}\ \bibnamefont {Chu}},
  \bibinfo {author} {\bibfnamefont {J.~K.}\ \bibnamefont {Jain}},\ and\
  \bibinfo {author} {\bibfnamefont {S.-Q.}\ \bibnamefont {Shen}},\ }\bibfield
  {title} {\bibinfo {title} {Topological anderson insulator},\ }\href
  {https://doi.org/10.1103/PhysRevLett.102.136806} {\bibfield  {journal}
  {\bibinfo  {journal} {Phys. Rev. Lett.}\ }\textbf {\bibinfo {volume} {102}},\
  \bibinfo {pages} {136806} (\bibinfo {year} {2009})}\BibitemShut {NoStop}%
\bibitem [{\citenamefont {Zhang}\ \emph {et~al.}(2012)\citenamefont {Zhang},
  \citenamefont {Chu}, \citenamefont {Zhang},\ and\ \citenamefont
  {Shen}}]{Zhang2012}%
  \BibitemOpen
  \bibfield  {author} {\bibinfo {author} {\bibfnamefont {Y.-Y.}\ \bibnamefont
  {Zhang}}, \bibinfo {author} {\bibfnamefont {R.-L.}\ \bibnamefont {Chu}},
  \bibinfo {author} {\bibfnamefont {F.-C.}\ \bibnamefont {Zhang}},\ and\
  \bibinfo {author} {\bibfnamefont {S.-Q.}\ \bibnamefont {Shen}},\ }\bibfield
  {title} {\bibinfo {title} {Localization and mobility gap in the topological
  anderson insulator},\ }\href {https://doi.org/10.1103/PhysRevB.85.035107}
  {\bibfield  {journal} {\bibinfo  {journal} {Phys. Rev. B}\ }\textbf {\bibinfo
  {volume} {85}},\ \bibinfo {pages} {035107} (\bibinfo {year}
  {2012})}\BibitemShut {NoStop}%
\bibitem [{\citenamefont {Meier}\ \emph {et~al.}(2018)\citenamefont {Meier},
  \citenamefont {An}, \citenamefont {Dauphin}, \citenamefont {Maffei},
  \citenamefont {Massignan}, \citenamefont {Hughes},\ and\ \citenamefont
  {Gadway}}]{Meier2018}%
  \BibitemOpen
  \bibfield  {author} {\bibinfo {author} {\bibfnamefont {E.~J.}\ \bibnamefont
  {Meier}}, \bibinfo {author} {\bibfnamefont {F.~A.}\ \bibnamefont {An}},
  \bibinfo {author} {\bibfnamefont {A.}~\bibnamefont {Dauphin}}, \bibinfo
  {author} {\bibfnamefont {M.}~\bibnamefont {Maffei}}, \bibinfo {author}
  {\bibfnamefont {P.}~\bibnamefont {Massignan}}, \bibinfo {author}
  {\bibfnamefont {T.~L.}\ \bibnamefont {Hughes}},\ and\ \bibinfo {author}
  {\bibfnamefont {B.}~\bibnamefont {Gadway}},\ }\bibfield  {title} {\bibinfo
  {title} {Observation of the topological anderson insulator in disordered
  atomic wires},\ }\href
  {https://www.science.org/doi/abs/10.1126/science.aat3406} {\bibfield
  {journal} {\bibinfo  {journal} {Science}\ }\textbf {\bibinfo {volume}
  {362}},\ \bibinfo {pages} {929} (\bibinfo {year} {2018})}\BibitemShut
  {NoStop}%
\bibitem [{\citenamefont {Bardeen}\ \emph {et~al.}(1957)\citenamefont
  {Bardeen}, \citenamefont {Cooper},\ and\ \citenamefont
  {Schrieffer}}]{Bardeen1957}%
  \BibitemOpen
  \bibfield  {author} {\bibinfo {author} {\bibfnamefont {J.}~\bibnamefont
  {Bardeen}}, \bibinfo {author} {\bibfnamefont {L.~N.}\ \bibnamefont
  {Cooper}},\ and\ \bibinfo {author} {\bibfnamefont {J.~R.}\ \bibnamefont
  {Schrieffer}},\ }\bibfield  {title} {\bibinfo {title} {Theory of
  superconductivity},\ }\href {https://doi.org/10.1103/PhysRev.108.1175}
  {\bibfield  {journal} {\bibinfo  {journal} {Phys. Rev.}\ }\textbf {\bibinfo
  {volume} {108}},\ \bibinfo {pages} {1175} (\bibinfo {year}
  {1957})}\BibitemShut {NoStop}%
\bibitem [{\citenamefont {Ryu}\ and\ \citenamefont
  {Hatsugai}(2002)}]{ryu2002topological}%
  \BibitemOpen
  \bibfield  {author} {\bibinfo {author} {\bibfnamefont {S.}~\bibnamefont
  {Ryu}}\ and\ \bibinfo {author} {\bibfnamefont {Y.}~\bibnamefont {Hatsugai}},\
  }\bibfield  {title} {\bibinfo {title} {Topological origin of zero-energy edge
  states in particle-hole symmetric systems},\ }\href
  {https://doi.org/10.1103/PhysRevLett.89.077002} {\bibfield  {journal}
  {\bibinfo  {journal} {Phys. Rev. Lett.}\ }\textbf {\bibinfo {volume} {89}},\
  \bibinfo {pages} {077002} (\bibinfo {year} {2002})}\BibitemShut {NoStop}%
\bibitem [{\citenamefont {Sun}\ \emph {et~al.}(2012)\citenamefont {Sun},
  \citenamefont {Liu}, \citenamefont {Hemmerich},\ and\ \citenamefont
  {Das~Sarma}}]{sun2012topological}%
  \BibitemOpen
  \bibfield  {author} {\bibinfo {author} {\bibfnamefont {K.}~\bibnamefont
  {Sun}}, \bibinfo {author} {\bibfnamefont {W.~V.}\ \bibnamefont {Liu}},
  \bibinfo {author} {\bibfnamefont {A.}~\bibnamefont {Hemmerich}},\ and\
  \bibinfo {author} {\bibfnamefont {S.}~\bibnamefont {Das~Sarma}},\ }\bibfield
  {title} {\bibinfo {title} {Topological semimetal in a fermionic optical
  lattice},\ }\href {https://www.nature.com/articles/nphys2134} {\bibfield
  {journal} {\bibinfo  {journal} {Nature Physics}\ }\textbf {\bibinfo {volume}
  {8}},\ \bibinfo {pages} {67} (\bibinfo {year} {2012})}\BibitemShut {NoStop}%
\bibitem [{\citenamefont {Matsuura}\ \emph {et~al.}(2013)\citenamefont
  {Matsuura}, \citenamefont {Chang}, \citenamefont {Schnyder},\ and\
  \citenamefont {Ryu}}]{matsuura2013protected}%
  \BibitemOpen
  \bibfield  {author} {\bibinfo {author} {\bibfnamefont {S.}~\bibnamefont
  {Matsuura}}, \bibinfo {author} {\bibfnamefont {P.-Y.}\ \bibnamefont {Chang}},
  \bibinfo {author} {\bibfnamefont {A.~P.}\ \bibnamefont {Schnyder}},\ and\
  \bibinfo {author} {\bibfnamefont {S.}~\bibnamefont {Ryu}},\ }\bibfield
  {title} {\bibinfo {title} {Protected boundary states in gapless topological
  phases},\ }\href
  {https://iopscience.iop.org/article/10.1088/1367-2630/15/6/065001} {\bibfield
   {journal} {\bibinfo  {journal} {New Journal of Physics}\ }\textbf {\bibinfo
  {volume} {15}},\ \bibinfo {pages} {065001} (\bibinfo {year}
  {2013})}\BibitemShut {NoStop}%
\bibitem [{\citenamefont {Wang}\ \emph {et~al.}(2017)\citenamefont {Wang},
  \citenamefont {Miao}, \citenamefont {Jin},\ and\ \citenamefont
  {Chen}}]{wang2017characterization}%
  \BibitemOpen
  \bibfield  {author} {\bibinfo {author} {\bibfnamefont {Y.}~\bibnamefont
  {Wang}}, \bibinfo {author} {\bibfnamefont {J.-J.}\ \bibnamefont {Miao}},
  \bibinfo {author} {\bibfnamefont {H.-K.}\ \bibnamefont {Jin}},\ and\ \bibinfo
  {author} {\bibfnamefont {S.}~\bibnamefont {Chen}},\ }\bibfield  {title}
  {\bibinfo {title} {Characterization of topological phases of dimerized kitaev
  chain via edge correlation functions},\ }\href
  {https://doi.org/10.1103/PhysRevB.96.205428} {\bibfield  {journal} {\bibinfo
  {journal} {Phys. Rev. B}\ }\textbf {\bibinfo {volume} {96}},\ \bibinfo
  {pages} {205428} (\bibinfo {year} {2017})}\BibitemShut {NoStop}%
\bibitem [{\citenamefont {Miao}\ \emph {et~al.}(2017)\citenamefont {Miao},
  \citenamefont {Jin}, \citenamefont {Zhang},\ and\ \citenamefont
  {Zhou}}]{miao2017exact}%
  \BibitemOpen
  \bibfield  {author} {\bibinfo {author} {\bibfnamefont {J.-J.}\ \bibnamefont
  {Miao}}, \bibinfo {author} {\bibfnamefont {H.-K.}\ \bibnamefont {Jin}},
  \bibinfo {author} {\bibfnamefont {F.-C.}\ \bibnamefont {Zhang}},\ and\
  \bibinfo {author} {\bibfnamefont {Y.}~\bibnamefont {Zhou}},\ }\bibfield
  {title} {\bibinfo {title} {Exact solution for the interacting kitaev chain at
  the symmetric point},\ }\href
  {https://doi.org/10.1103/PhysRevLett.118.267701} {\bibfield  {journal}
  {\bibinfo  {journal} {Phys. Rev. Lett.}\ }\textbf {\bibinfo {volume} {118}},\
  \bibinfo {pages} {267701} (\bibinfo {year} {2017})}\BibitemShut {NoStop}%
\bibitem [{\citenamefont {Young}\ and\ \citenamefont
  {Rieger}(1996)}]{young1996numerical}%
  \BibitemOpen
  \bibfield  {author} {\bibinfo {author} {\bibfnamefont {A.~P.}\ \bibnamefont
  {Young}}\ and\ \bibinfo {author} {\bibfnamefont {H.}~\bibnamefont {Rieger}},\
  }\bibfield  {title} {\bibinfo {title} {Numerical study of the random
  transverse-field ising spin chain},\ }\href
  {https://doi.org/10.1103/PhysRevB.53.8486} {\bibfield  {journal} {\bibinfo
  {journal} {Phys. Rev. B}\ }\textbf {\bibinfo {volume} {53}},\ \bibinfo
  {pages} {8486} (\bibinfo {year} {1996})}\BibitemShut {NoStop}%
\end{thebibliography}

%

\end{document}